\documentclass[12pt,preprint]{aastex}
\slugcomment{Accepted and scheduled for publication  
in {\it The Astrophysical Journal},  for the ApJ April 20, 2013, v 767, 2nd issue}  
\def\lax {\ifmmode{_<\atop^{\sim}}\else{${_<\atop^{\sim}}$}\fi}  
\def\gax {\ifmmode{_>\atop^{\sim}}\else{${_>\atop^{\sim}}$}\fi}  
\def\gtorder{\mathrel{\raise.3ex\hbox{$>$}\mkern-14mu
             \lower0.6ex\hbox{$\sim$}}}

\def\qcor{Q_{\rm cor}}
\def\qtot{Q_{\rm tot}}

\def\qd{Q_{\rm disk}}

\def\cm2{cm$^{-2}$}
\def\s1{s$^{-1}$}

\def\kte{kT_{\rm e}}

\begin{document}

\title{Spectral state evolution of 4U~1820-30: the stability of the spectral index of Comptonization tail} 





\author{Lev Titarchuk\altaffilmark{1}, Elena Seifina\altaffilmark{2} and Filippo Frontera\altaffilmark{3}  }

\altaffiltext{1}{Dipartimento di Fisica, Universit\`a di Ferrara, Via Saragat 1, I-44122 Ferrara, Italy, email:titarchuk@fe.infn.it; George Mason University Fairfax, VA 22030;   
Goddard Space Flight Center, NASA,  code 663, Greenbelt  
MD 20770, USA; email:lev@milkyway.gsfc.nasa.gov, USA}
\altaffiltext{2}{Moscow M.V.~Lomonosov State University/Sternberg Astronomical Institute, Universitetsky 
Prospect 13, Moscow, 119992, Russia; seif@sai.msu.ru}
\altaffiltext{3}{Dipartimento di Fisica, Universit\`a di Ferrara, Via Saragat 1, I-44122  Ferrara, Italy, email:frontera@fe.infn.it
}

\begin{abstract}
We analyze the X-ray spectra and their timing properties 
of the  compact 
X-ray binary 
 4U~1820-30.
We establish 
spectral transitions 
in this source  seen  with  {\it Beppo}SAX and the {\it Rossi} X-ray Timing Explorer
({\it RXTE}). 
During the {\it RXTE} observations (1996 -- 2009), the source  were approximately $\sim75\%$ of 
 its time 
 in the soft state making  the {\it lower banana} and {\it upper banana} 
 transitions combined with long-term  {\it low-high state} transitions. 
 We reveal 
 that all of the X-ray spectra  
 of 4U~1820-30
are  fit 
by a composition  of a thermal (blackbody) component, a Comptonization
component (COMPTB) and 
a {\it Gaussian-}line component. Thus using this spectral analysis  we find 
that the photon power-law index $\Gamma$ of the Comptonization component is almost
unchangeable 
($\Gamma\sim 2$) while 
the electron temperature $kT_e$ changes from 2.9 to 21 keV
during these spectral events.
We also establish  that for  these spectral events  the  normalization of {\it COMPTB}  component 
(which is proportional to mass accretion rate $\dot M$) increases by factor 8
when $kT_e$ decreases from 21 keV to 2.9 keV.  
Before this index stability effect was also  found analyzing  X-ray data 
for  {\it Z-}source GX~340+0 and for {\it atolls}, 4U~1728-34,  GX~3+1. 
Thus, we can {\it suggest}
that this spectral stability property is  a spectral signature of an accreting  neutron star source.  On the other  hand
%
in a black hole binary 
$\Gamma$ monotonically increases with $\dot M$ 
and ultimately   its value saturates at  large $\dot M$.  

\end{abstract}


\section{Introduction}

Accreting neutron stars (NS) can be  observationally classified using a color-color diagram (CCD) 
into  two distinct categories, 
{\it atoll} and {\it Z}  sources,
based on their different CCD forms 
when the source undergoes the spectral and luminosity  changes.
Along with this phenomenological difference  between {\it atolls} and {\it Z-}sources in terms of  the CCD there are   important X-ray spectral and timing characteristics, which are essentially  different for these types of NS  low mass X-ray binaries (LMXB). 
The main observational  difference between these types  is the specific range of luminosity changes. The 
{\it atolls} are observed
when their luminosity changes 
from 0.01 up to 0.5 of the Eddington limit $L_{\rm Edd}$ (see Christian \& Swank 1997; Ford et al. 2000) 
 while {\it Z-}sources are seen 
when the resulting luminosity is near Eddington regime (e.g. Seifina, Titarchuk \& Frontera 2013).
In this Paper we present our analysis of {\it peculiar atoll}  4U~1820-30, which is a {\it bright atoll} 
source in terms of  its luminosity and, at the same time,  it is a {\it typical atoll} source in terms 
of timing evolution. On the other hand, during bright phase 4U~1820-30 is as bright as  a
subclass of persistent bright {\it atolls}  (GX~13+1,GX~9+1, GX~9+9 and GX~3+1), 
but 4U~1820-30 shows larger range of luminosity and demonstrates all states in terms of CCD
(from the $island$ to $banana$ states), whereas these aforementioned  bright {\it atolls} are only seen 
 in the banana state  (e.g. Hasinger \& van der Klis 1989). 

Furthermore,  4U~1820-30 has a maximal luminosity about  0.5 of $L_{\rm Edd}$ and thus it 
adjoins to {\it Z} source luminosity range. 
Also  4U~1820-30 shows Type-I  X-ray bursts and characteristic timing features 
as  a typical {\it atoll} 
demonstrating an evolution of band-limited noise (BLN), very low frequency noise (VLFN) 
components and low frequency quasi-periodic oscillations (LFQPOs) (in  20-40 Hz  range)
 when
 it evolves from the {\it banana} to {\it island} states. However, opposite to 
the other  {\it atolls}, 4U~1820-30 exhibits LFQPOs with frequencies near 6 Hz  during the {\it banana} state 
(Wijnands, van der Klis, \& Rijkhorst 1999; also see Sect.~6.2 in this Paper), which are usually seen
in {\it Z-}sources during the Normal branch. Thus, 4U~1820-30 combines 
properties of ordinary  and bright {\it atolls} and {\it Z-}sources in terms of timing  and spectral evolution, 
luminosity and detection  of 6 Hz QPO. 


4U~1820-30 
is a 
LMXB 
observed at 0.66'' from the center 
of the NGC 6624
cluster. \cite{Grindlay76} were the first who  identified  this source as a  Type-I X-ray burst. 
Kuulkers et al. (2003) estimated  the distance $d=7.6\pm 0.4$ kpc to 4U~1820-30 
assuming that the peak luminosity equals to 
$L_{\rm Edd}$ for
He burst atmosphere. Vacca, Lewin \& van Paradijs (1986)  estimate 
the distance to 4U~1820-30 as 6.4$\pm$0.6 kpc using the analysis of UV diagrams for  NGC 6624.  
\cite{rap87} find that   the binary system comprises of a He white dwarf of mass of 0.06$\pm$0.08 M$_{\odot}$ 
and NS, [with a mass later  evaluated by Shaposhnikov \& Titarchuk (2004) as 
$\sim$1.3 M$_{\odot}$],  orbiting at the period  
of 11.4 minutes \citep{stel87}. 

\cite{hasinger89}   classify 
4U~1820-30 
as an{ \it atoll} source.   
\cite{PT84}, \cite{Simon03} and \cite{Wen06} find that the flux  varies between the soft and  the hard
states  ({\it banana} and  {\it island}  ones respectively) are quasi-periodic with period at $\sim$170 d and  these flux variations  have been  suggested  related to  tidal effects of a 
remote third star [\cite{Chou01} and \cite{Zdziarski07}]. 
{\it  RXTE} observations revealed 4U~1820-30 as a prominent
source of kilohertz quasi-periodic oscillations (kHz QPO)  (Smale 1997). 
X-ray bursts are only  observed 
at low fluxes  [e.g. \cite{clark77}].  Furthermore, \cite{Cornelisse03} and \cite{Zhang98}  find  that 
the observed kHz QPOs correlate with 
the flux which probably  suggests 
that these variations
are  because of  a luminosity  change caused by  change of mass accretion rate. 
 
Strohmayer \& Bildsten (2004) established  
that the short ($\sim$10 -- 15 s) Type-I outbursts  are because of 
the unstable thermonuclear burning of  mixture of hydrogen and helium 
at  the  NS atmosphere bottom. 
SAS-3 observations
showed strong evidence that X-ray bursts can only occur    in its low-intensity state (Clark et al. 1977). 
All bursts observed from 4U~1820-30 indicate that  a photosphere expends  with an increase of  
a photospheric radius 
by a factor of 20. Such an expansion leads to strong softening of the resulting spectrum
(see e.g. Strohmayer \& Bildsten 2004).
Moreover, a several hour long "superbursts" was observed from 4U 1820-30 on September 9, 1999. It
is now  understood 
that superbursts can be caused by the burning in the carbon ashes produced by
Type-I bursts (Strohmayer \& Brown 2002).

{\it Einstein}, EXOSAT, $Ginga$, ASCA,  and $Beppo$\-SAX  also observed 4U~1820-30.  
 Many spectral models have been applied to fit the 
observed X-ray spectra. For example, 
models which are a sum  of a blackbody (BB)  with 
thermal bremss\-trahlung    
or  a blackbody with a power law combined with exponential
cutoff (CPL). 
A more detailed model were developed  using a  Comptonization spectrum 
by Sunyaev \& Titarchuk (1980)  (see CompST model in XSPEC) combined with a blackbody.  
Note, White et al. (1986) and Christian \& Swank 1997) show  that the models based   on thermal bremsstrahlung  mechanism  are unphysical.  The  emission measures found using this model have to be  of order  $\sim 10^{60}$~ cm$^{-3}$  and they are too large  for the plasma cloud  near NS which 
radius is  only of order of $10^{6}$ cm. 

Therefore 
the CPL and CompST components combined  with  a blackbody were applied to fit  X-ray spectra of 4U~1820-30
[see e.g. \cite{Bloser00}]. 
The BB temperature $kT_{BB}$, the 
photon index $\Gamma$ and exponential  
cutoff energy $E_C$ are the CPL+BB model parameters.   \cite{Bloser00} also included  
photoelectric absorption at low energies 
 using the cross section of Morrison \& McCamman (1983).  
 White et al. 1986 and Hirano et al.
(1987)  show that 
a  {\it Gaussian} at $\sim$6.7 keV is often needed to take into account a blend of $K_{\alpha}$ 
 iron lines.
Parsignault \& Grindlay (1978) applying a power-law fit to   the 4U~1820-30 ANS data (ANS is an abbreviation  of the Astronomische Nederlandse Satelliet)    found
X-ray  spectral  changes due intensity variations.
They obtain  that the
photon index $\Gamma$  changes from 2 
at high count rates to 1.4 when count rate is low. In other words, they find that spectrum becomes harder when luminosity decreases.
Stella, White, \& Priedhorsky (1987) used  the CPL+BB 
model  to fit the data of the EXOSAT ME instrument in the energy range from  1 to 30 keV. These particular EXOSAT data  were  obtained 
during 1984 -- 1985. The source was found at a $high$ luminosity  state 
(6.0$\times$10$^{37}$ erg s$^{-1}$) and a $low$ luminosity  state  (2.0$\times$10$^{37}$ erg s$^{-1}$). The best-fit parameters of these EXOSAT spectra $\Gamma$,  $E_C$ and $kT_{BB}$  are at 1.7, 12 keV and  2 keV respectively in the high state while   $\Gamma\sim $2.5, $E_C>$30 keV, and  $kT_{BB}$=2.3 keV in  the low state.  

Smale et al. (1994) analyzed  the ASCA/GIS data
for 4U 1820-30  which  was observed 
 in the low state  in 1993. They  
 fit  the 0.6-11 keV spectrum using 
 the CompST + BB model. 
 The best-fit CompST parameters  were  around 3.6, 13.5 and 0.76 keV for the plasma temperature, optical
 and a blackbody temperature
 respectively.
Christian \& Swank (1997) reported on 
the $Einstein$ (SSS + MPC) 1978 observation in the 0.5 --  20 keV energy range. They found the source was in the high state characterized by the  luminosity 
of  5.5$\times$10$^{37}$ erg s$^{-1}$ and 
the best-fit parameters of    the CompST+BB model were very similar to those  obtained using the ASCA data. 

Piraino et al. (1999) and  Kaaret et al. (1999) analyzed  the first observations of 4U~1820-30 
extended  above $\sim$30 keV. For this analysis they used the observations implemented by the NFIs (0.1 -- 200 keV) of $Beppo$SAX in 1998.
The best-fit of  the observed spectrum in the 0.3 -- 40 keV energy  range   give $kT_{BB}$=0.47 keV,   
$\Gamma$=0.55, and $E_C$=4.5 keV for the CPL+BB model , whereas  the best-fit parameters of the CompST+BB model are $kT_{BB}$=0.46 -- 0.66 keV,  $kT_e$=2.83 keV and $\tau$=13.7.
It is worth noting that  the instruments with a response below 1 keV provide  low values of $kT_{BB}$. 
Although, the  range of luminosities and the best-fit parameters  inferred using $Beppo$SAX and  the CompST + BB model gave very similar values with respect to that  obtained using 
the other instruments (see above).  
%

The $Beppo$SAX Phoswich Detection System (PDS) could not  detect
the emission from  4U~1820-30 above 40 keV.  Note,  BATSE [see Bloser et al. 1996] also failed to find this source  in the 20 -- 100 keV range  
during the first four years of the CGRO. 
For the first time a high energy tail above 50 keV has been found  by INTEGRAL 
in the hard state of 4U~1820-30 \citep{tarana06}, which put this  
source in the list 
of X-ray bursters which exhibit  high-energy emission.


It is interesting to note that also other {\it atolls} can be described by similar spectral models. 
\cite{LRH07}, hereafter LRH07,  pointed out that  at the low-L$_X$ end of the soft-state track a weak Comptonization component is needed. 
LRH07 modified the BPL and COMPTT models  applying their modifications to atolls  Aql~X-1 and 4U~1608-52. They were interested to find out  how much energy 
is directly visible as a pure thermal radiation and thus one can obtain the remaining fraction for the Comptonized radiation $f$.  In this sense this approach is similar to that  using  the COMPTB model (see below \S 3) .    
In this way, to account 
for specific spectra
LRH07 describe  the hard state by a BB+BPL model and the soft state by means of a three component model, MCD+BB+CBPL, where 
 CBPL is a broken power law with the high energy cutoff 
 taking into account  the Comptonization effect. 


In this Paper  we show   a thorough  X-ray spectral-timing analysis of the data for 4U~1820-30 using 
 the {\it Beppo}SAX  and 
{\it RXTE}/PCA/HEXTE available  observations  which were made  during  
1998 -- 1999  and 1996 - 2009 years respectively.    Unlike the past spectral analysis, we adopt an unified model, capable describing the spectra observed during both the soft and hard states.  
The full  list of observations used in our data analysis is present  in \S 2 and Tables 1 and 2  while 
we describe, in detail  our spectral model  and  spectral analysis using this model  in \S 3.
We 
interpret 
X-ray spectral-timing  evolution   when the source undergoes  the spectral state transition in \S\S 4$-$6.  We explain our results in detail  and come  to the final  conclusions in \S\S 7$-$8.

\section{Data Selection \label{data}}

We obtain broad band energy spectra of the source using 
data from  three {\it Beppo}SAX Narrow
Field Instruments (NFIs), namely  the Low Energy Concentrator
Spectrometer (LECS) with the 0.3$-$4 keV energy band,  
the Medium Energy Concentrator Spectrometer
(MECS) with the  1.8$-$10 keV band and the Phoswich Detection
System (PDS) with  the 15-200 keV band 
[see  \cite{parmar97}; \cite{boel97};  \cite{fron97} respectively].

  We use the SAXDAS data analysis package for  data processing. 
 We renormalized the LECS data based on the MECS data. We treat  relative normalizations of the NFIs 
as free parameters when we proceed with 
 model fitting, but  we fix 
 the MECS normalization at 1.
Each of these normalizations is  controlled if they
 are in a standard range for a given
 instrument
\footnote{http://heasarc.nasa.gov/docs/sax/abc/saxabc/saxabc.html}.
Furthermore,  we  rebinned   the spectra to obtain 
significant data points.  The LECS spectra are rebinned using a binning factor 
which varies with
energy
(Sect. 3.1.6 of Cookbook for the BeppoSAX NFI spectral analysis) implementing 
 rebinning template files
 in GRPPHA of
 XSPEC \footnote{http://heasarc.gsfc.nasa.gov/FTP/sax/cal/responses/grouping}. 
 The PDS spectra are rebinned with a  linear binning
 factor 2,  namely we group two bins together leading to the 
 bin width of 1 keV.  For all of these spectra we use  a systematic error of 1\%. 
The {\it Beppo}SAX observations implemented in our analysis  are shown in Table 1.

We   also use publicly available the {\it RXTE}  data sets 
\citep{bradt93}  which were  obtained from April 1997 to March 2009. 
In total, they  include 92 observations taken at
different states of the source.
The LHEASOFT/FTOOLS
5.3 software package were applied to process the data.
Also for our spectral analysis 
we apply PCA {\it Standard 2} mode data, collected 
in the 3 -- 20~keV energy range and the most recent release of PCA response 
calibration (ftool pcarmf v11.1).
We use the  standard dead time correction  to the data. 

A background corrected  in  off-source observations is subtracted from the data.
We use   only data  from 20  to  150~keV energy  
in order to avoid 
the problems related to  the HEXTE response and 
background determination. 
 We apply 
 the GSFC public archive to analyze all available data sets 
(see http://heasarc.gsfc.nasa.gov).  We present a full list  of  observations covering 
the source evolution during  different spectral  state events in  Table 2. 


We implement  an analysis of  {\it thirteen} years {\it RXTE} observations  of 4U~1820-30   
for 7 intervals  (see  blue rectangles  in Figure 1 in \cite{tsf13}, hereafter TSF13).
We fitted the {\it RXTE} 
energy spectra
using XSPEC astrophysical fitting software. 
For our data analysis we have also applied the public available  4U~1820-30 data in the energy range from 2 to 10 keV from the  All-Sky Monitor (ASM/{\it RXTE})  for  all observation scans.

According to ASM monitoring system 4U~1820-30 shows long-term variations with possible period $\sim$176 days  of the 2 -- 10 keV flux  [see Fig. 1 in TSF13
and  \cite{PT84}; \cite{Simon03} and \cite{Wen06}]. 
The count rate changes in the interval  of 5$-$35 counts s$^{-1}$  throughout each cycle 
(see in Fig. 1 in  TSF13).
Our {\it RXTE} spectral studies are directed to investigate: 
i) the continuum spectrum, in particular, 
the hard X-ray tail  and its  evolution during  long-term flux variations, 
ii) the variation ($\gax$10 sec) of the best-fit spectral parameters for  short- and long-term phases, and 
iii) the dependence of the spectral index and the electron temperature on the total flux 
and accretion rate. Data from the PCA and HEXTE detectors as well as $Beppo$SAX detectors have been used 
to constrain spectral fits, 
while ASM data provided long-term intensity state monitoring. Results of our long-term study 
of 4U~1820-30 are present, in detail, in the next sections and compared with our previous results for 
4U~1728-34 and GX~3+1.

We use the broadband energy spectra 
of 
{\it Beppo}SAX (Boella et al. 1997) and {\it RXTE}  (Bradt, Rothschild, \& Swank 1993)  combined with the high-timing resolution of {\it RXTE}  to 
 study  
 short and long term  spectral and timing evolution of {\it atoll} 
 sources. 


\section{Spectral Analysis \label{spectral analysis}}

Unlike the past analyses of the source spectral data discussed in the Introduction, in our study
we make use of an unified  model for both soft and hard states. In this way we 
have an opportunity to compare X-ray spectra of 4U~1820-30 in all states.

 In our spectral model,  we use an assumption, that the accretion material passes 
through the accretion disk [for example, through the standard Shakura-Sunyaev 
disk \citep{ss73}] and  the  transition layer (TL) \citep{tlm98} where soft photons coming from 
the disk and NS surface 
%
are  Comptonized 
off hot plasma 
(see also Fig.~2 in ST12). 
The Earth  observer can also observe  directly  some fraction of these disk and NS seed  photons.

Thus, 
{our  input model} 
is a  sum of  Comptonization component
 ($COMPTB$), 
which is the XSPEC Contributed model\footnote{http://heasarc.gsfc.nasa.gov/docs/software/lheasoft/xanadu/xspec/models/comptb.html},
[see \cite{F08}, hereafter F08] and soft {\it blackbody}  and  line ($Gaussian$)  components. 
  The parameters of the $COMPTB$  component are
the seed photon temperature
$kT_s$, the electron (plasma) temperature $kT_e$, the energy index  $\alpha$ ($=\Gamma-1$) of the Comptonization spectrum,   the illumination
fraction of the Comptonized region (TL), $f$ [$f=A/(1+A)$]
and the normalization of 
the seed  photons illuminating the  Comptonized region,  $N_{COMPTB}$. 
We   include a  {\it Gaussian} component  in the model characterized by the parameters
$E_{line}$, 
$\sigma_{line}$  
$N_{line}$ which are a centroid line energy,  the line width and the normalization correspondingly. 
We also include a $blackbody$ component  and the interstellar absorption  in our model  
characterized by the 
parameters: the normalization $N_{BB}$,  the color  temperature $T_{BB}$ and 
a column density $N_H$ respectively.

We  fix  the index of the seed photon spectrum  $\alpha=2$ (or $\gamma= \alpha+1=3$).
Namely, we suggest that  this seed photon spectrum is a  blackbody-like.
We neglect 
 the  bulk inflow effect with respect to  the  thermal Comptonization  assuming that a bulk parameter
 $\delta=0$. 
The   parameter $\log(A)$ of  the $COMPTB$ component   is fixed  at 2 because  the best-fit 
$\log(A)\gg1$. Then   
 $f=A/(1+A)$ as the illumination fraction  parameter is  approximately 1, for any  $\log(A)\gg 1$.
We use a value of 
$N_H=3.00\times 10^{21}$ cm$^{-2}$ estimated  by~\cite{Bloser00} for 4U 1820-30. 
 We find satisfactory 
fits  using  our 
model for both  {\it Beppo}SAX  and {\it RXTE}  observations of 4U~1820-30
for all available data sets.

\subsection{{\it Beppo}SAX data analysis}

\begin{figure}[ptbptbptb]
\includegraphics[scale=1.0,angle=0]{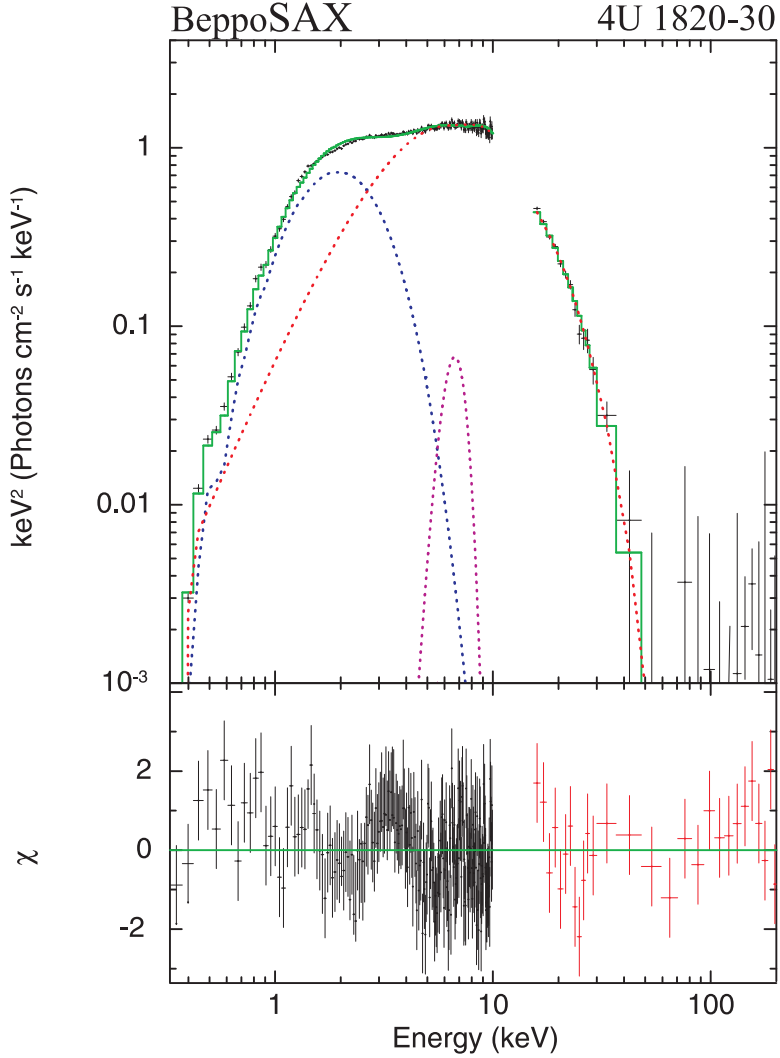}
\caption{$Top:$  the best-fit spectrum of 4U~1820-30 during {\it banana branch} events 
 in $E*F(E)$ units using {\it Beppo}SAX observation  20105004 carried out on 2 October 1997.   
The data are presented by crosses and the best-fit spectral  model   {\it wabs*(blackbody+Comptb+Gaussian)} by green line. 
The model components  are shown by  blue, red and crimson lines for {\it blackbdody}, {\it Comptb}  
and {\it Gaussian} components respectively. 
 $Bottom~panel$: 
 $\Delta \chi$ (reduced $\chi^2$=1.11 for 364 d.o.f).
 The best-fit model parameters are $\Gamma=2.00\pm0.04$, $kT_e=3.25\pm 0.02$ keV, $E_{line}=6.7\pm 0.1$ keV
 (see more details in Table 3).
}
\label{BeppoSAX_spectra}
\end{figure}

Table 3 shows the data analysis results
for the broad-band {\it Beppo}SAX spectra. 
On the $top$ of Figure~\ref{BeppoSAX_spectra} we {present  an example of  
the
 {\it Beppo}SAX
spectrum along with  its   best-fit using  our   model while}  
in the 
{\it bottom } panel we demonstrate 
 $\Delta \chi$
 (reduced $\chi^2$=1.11 for 364 d.o.f). 
 The   line emission is clearly  centered around 6.7 keV.
 We find   that  the width of this line of 0.8 keV  is  quite large and it is much wider than 
the instrumental response which width is smaller than 0.02 keV
\footnote{See ftp://heasarc.gsfc.nasa.gov/sax/cal/responses/98\_11}.
This  broad emission line at 6.7 keV can be a result of illumination of  highly ionized iron by X-ray 
continuum. 
Piraino et al. (2000)  suggest  that  
this broad line originates either  in 
an ionized innermost disk region or 
in  a hot corona above the disk.
A combination of two absorption edges related to  ionized iron, instead of a Gaussian line, can also describe  this part of the spectrum (D'Ai et al. 2006).   
The   $Laor$ relativistically  smeared line  or  reflection models   can be also  used to describe this line feature
(see Ng et al. 2010 and Egron et al. 2011 respectively).
However,  Seifina \& Titarchuk (2011) demonstrate that  
the model, wabs$\ast$(blackbody+COMPTB+Gaussian), which includes a a $Gaussian$ iron line, 
 can successfuly  fit the data  for  extensive   {\it RXTE} and $Beppo$SAX observations of 4U 1728-34.

We can interpret  this broad iron line detected in  4U~1820-30 in terms of  reprocessing 
 emission by a disk.  In addition, we  fit a smeared absorption edge  in the  7$-$8 keV range  using the $smedge$ XSPEC model (see Ebisawa et al. 1994). 
The edge energy 
is 7.7$\pm$0.5 keV for the $wabs*(Bbody+Comptb+Gaussian)*smedge$ model which  indicates to  the presence   of ionized material in the emission region. The smearing width is fixed at 10 keV. We include this  edge component  in the fits for all BeppoSAX  data (see
Table 3).

 We obtain that 
 $\alpha=1.03\pm0.04$ 
(or  $\Gamma=\alpha+1=2.03\pm0.04$)  
 for  all analyzed  {\it Beppo}SAX data,
  the  seed photon temperature $kT_s$  of the $COMPTB$ component is low variable 
  and its value is around 1.3 keV  whereas
$kT_{BB}$  varies in the interval from  0.58 to 0.69 keV (see  Tables 1, 3 for details). 

\subsection{{\it RXTE} data analysis}

  For all {\it RXTE}  fits  we  fix the   {\it blackbody} temperature 
$kT_{BB}=$0.7 keV which is  an upper limit  of that  in  our   analysis  of   the {\it Beppo}SAX data (see Table 3),
because {\it RXTE} detectors cannot give us  
reliable spectra  below 3 keV. 
In Table 4 we show the  best-fit  parameters   of  the  {\it RXTE}  spectra using our model.
 It is important to point out that  for {\it all  {\it RXTE} observations} of 4U 1820-30   the photon   index 
 $\Gamma$   only slightly changes  around 2
 ($\Gamma=1.99\pm 0.02$) while the best-fit  
 $kT_e$ 
 varies in the 2.5$-$21 keV range.

However the determination of the iron line profile 
 using the {\it RXTE} data  is a difficult problem caused by  
  the low-energy resolution of  PCA/{\it RXTE} detector. 
  Moreover, the inclusion of  a $smedge$ compoment in the spectral model for {\it RXTE} data 
does not improve fit quality any more. Therefore we apply our spectral model to {\it RXTE} data 
using a simple $Gaussian$ as the line component without $smedge$ modeling. 
The line width $\sigma_{line}$ 
does not  vary much and it is  always in the interval from  0.9 to 1.3 keV  during all spectral transitions. 
Therefore  we fix $\sigma_{line}$ 
at 
1.2 keV for all spectra during fitting procedure.
The values of the best-fit seed photon temperatures, $kT_{s}=1-1.3$ keV  
are consistent with that  obtained using 
 the {\it Beppo}SAX data 
 (see Table 3).



\begin{figure}[ptbptbptb]
\includegraphics[scale=1.00,angle=0]{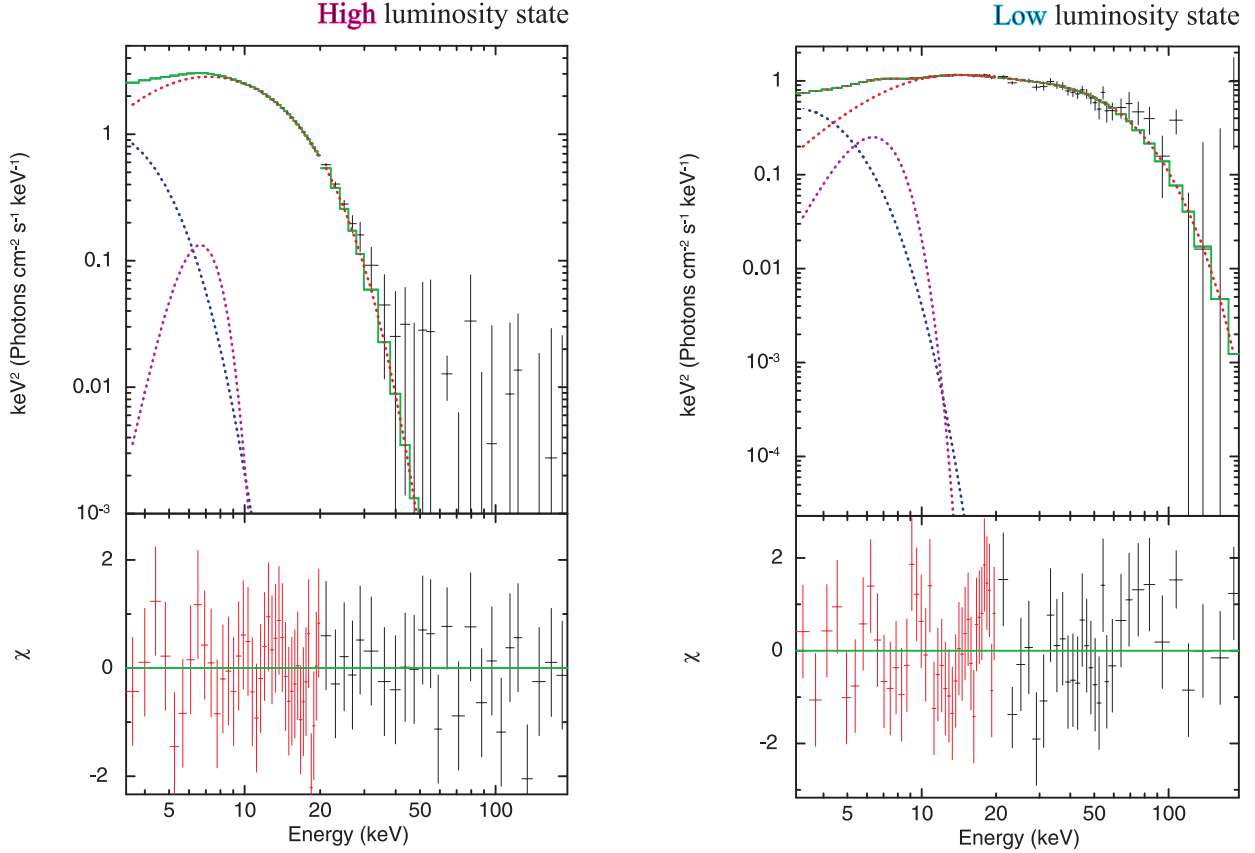}
\caption{Examples of typical $E*F(E)$ spectral diagram  of 4U~1820-30 during soft ({\it left panel}) and hard ({\it right panel}) state events.
The best-fit  {\it RXTE} spectra ($top$ panels) using the model $wabs*(Bbody+CompTB+Gaussian)$  with 
$\Delta\chi$ ($bottom$ panels) for the high-luminosity ($banana$) state [40017-01-11-00 observation, 
 $\chi^2_{red}$=1.00 for 78 d.o.f, {\it left panel}]  and for the low-luminosity ($island$) state [94090-01-04-00 observation, 
 $\chi^2_{red}$=1.10 for 78 d.o.f, {\it right panel}].
The best-fit model parameters are  
$\Gamma$=1.99$\pm$0.02, $kT_e$=2.94$\pm$0.01 keV and $E_{Gauss}$=6.53$\pm$0.06 keV (for the $soft$ state)
and   $\Gamma$=2.00$\pm$0.04, $kT_e$=12.54$\pm$0.09 keV and $E_{Gauss}$=6.35$\pm$0.04 keV (for the $hard$ state)    
(see more details  in Table 4). Blue, red and violet lines stand for 
$Bbody$, $CompTB$ and $Gauss$ components respectively.
}
\label{rxte_hard_soft_state_spectrum}
\end{figure}

In  Figure~\ref{rxte_hard_soft_state_spectrum} we show 
the representative examples of $E*F(E)$ spectral diagrams  of 4U~1820-30 during soft ({\it left panel}) 
and hard ({\it right panel}) state events.
The best-fit {\it RXTE} spectra ($top$ panels) in the model $wabs*(Bbody+CompTB+Gaussian)$  with 
$\Delta\chi$ ($bottom$ panels) for the  high-luminosity ($banana$) state [40017-01-11-00 observation, 
 $\chi^2_{red}$=1.00 for 78 d.o.f, {\it left panel}]  and for the low-luminosity ($island$) state [94090-01-04-00 observation, 
 $\chi^2_{red}$=1.10 for 78 d.o.f, {\it right panel}].
 The model best-fit 
parameters are  
$\Gamma=1.99\pm0.02$, $kT_e=2.94\pm0.01$ keV and $E_{Gauss}=6.53\pm0.06$ keV for the
 {\it soft} state;   
 $\Gamma=2.00\pm0.04$, $kT_e=12.54\pm0.09$ keV and  $E_{Gauss}=6.35\pm0.04$ keV  for 
 the {\it hard} state 
(see Table 4 for details). Violet,  blue, red  lines correspond to the $Gauss$,  
$Blackbody$ and $CompTB$  components, respectively.

\begin{figure}[ptbptbptb]
\includegraphics[scale=1.0, angle=0]{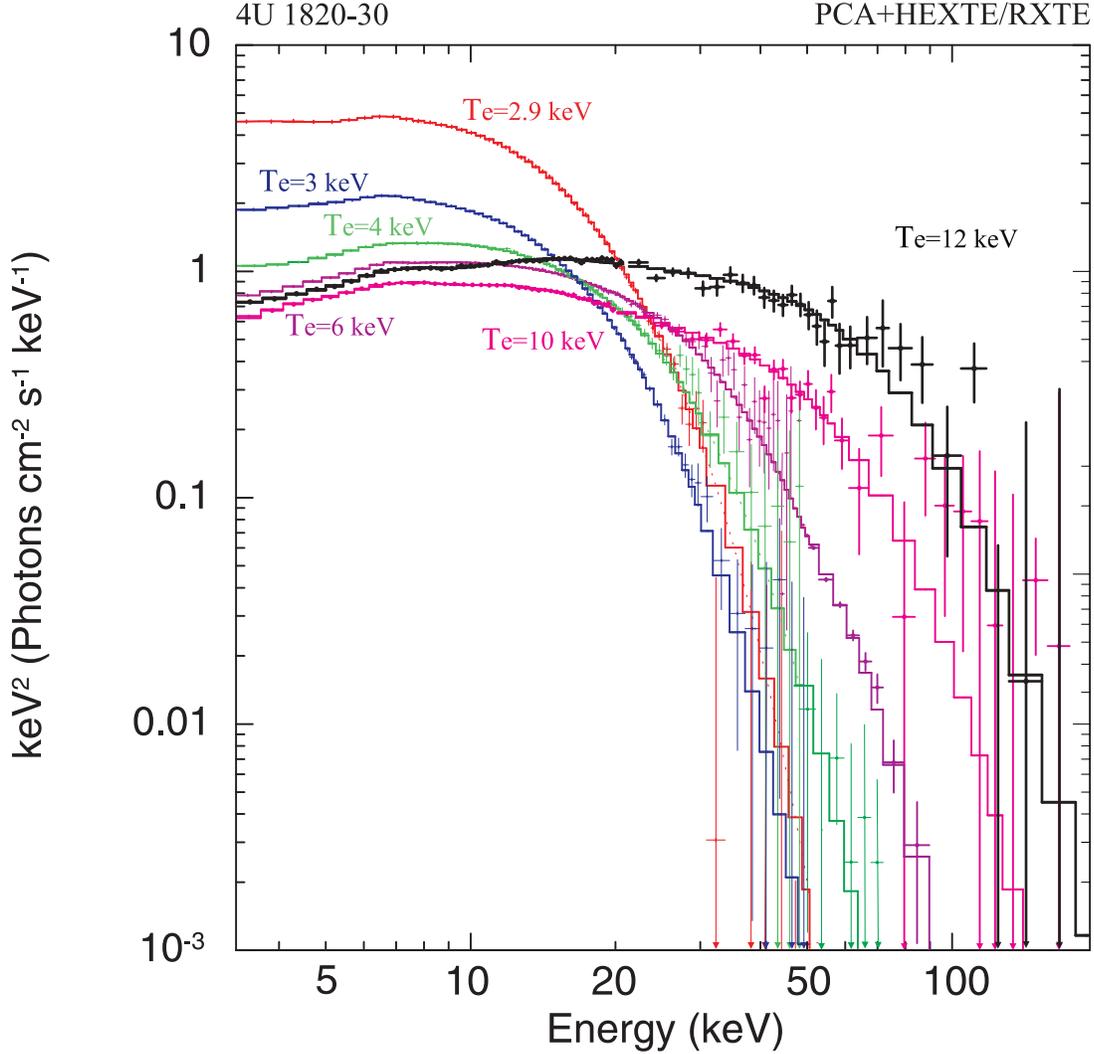}
\caption{Six representative $EF_E$ spectral diagrams which are related to different electron temperatures of TL
[$kT_e=$2.9 keV ($red$), 3 keV ($blue$), 4 keV ($green$), 6 keV ($voilet$), 10 keV ($pink$) and  12 keV ($black$)] using the model $wabs*(Blackbody+COMPTB+Gaussian)$ 
for $island-banana$ state transitions of 4U~1820-30. The  data are taken from {\it RXTE} observations 
30057-01-04-01 ($red$), 70030-03-07-020 ($blue$), 70030-03-05-02 ($green$), 70030-03-05-01 ($violet$), 40017-01-24-00 ($pink$) 
and 94090-01-04-00 ($black$).
}
\label{sp_compar}
\end{figure}

In Figure \ref{sp_compar} to illustrate  the spectral evolution 4U~1820-30  we show six representative $EF_E$ spectral diagrams for  different electron temperatures of  a Compton cloud
[$kT_e=$2.9 keV ($red$), 3 keV ($blue$), 4 keV ($green$), 6 keV ($violet$), 10 keV ($pink$) and  12 keV ($black$)] applying  the  $wabs*(Blackbody+COMPTB+Gaussian)$ model  
during $island-banana$ state transitions.

We show how the TL electron temperature 
$kT_e$ anti-correlates with the normalization $N_{\rm COMPTB}$
(proportional to $\dot M$) 
in Figure  \ref{norm_T_e}. For a comparison we add the  points for  4U 1728-34 and GX 3+1 [see  \cite{st11} and \cite{st12} respectively].  The electron temperature $kT_e$  decreases and reaches a saturation about  3 keV when mass accretion rate increases [see an explanation of this effect in \cite{ft11}, hereafter FT11].  

Our spectral model applied to the spectral data of {\it Beppo}SAX and {\it RXTE} shows a robust  performance throughout
all data sets.
Namely, a value of reduced
$\chi^2_{red}=\chi^2/N_{d.o.f}$, where $N_{d.o.f}$ is a number of degree of freedom, 
is  
around 1.0 for most of observations. 
$\chi^2_{red}$ is about  1.5 for less than  3\%
of spectra with high counting statistics  but  $\chi^2_{red}$ is never above  a rejection 
limit of 1.6. 
Note the high residuals of  the  poor fit  
spectra  (2 among 234 spectra for which $\chi^2$=1.55)   
occur in  the iron line region.  
As was shown by $Beppo$SAX analysis 
  a shape of the iron line is more complex than a simple Gaussian 
(see  discussion in \S  3.1). 
Probably, the fits of this line indicate to  a broad line,  
which shape and width   can be a result of   scattering of the  line photons in the hot plasma (TL) 
along  with iron $smedge$ effect there. 
However, we cannot resolve this line 
complexity  using
the {\it RXTE} data. 


Thus   using broad band 
{\it Beppo}SAX observations we can obtain  the best-fit parameters of our spectral model 
whereas
due to   large-time coverage  
of 4U 1820-30 by {\it RXTE} we  are capable to  
study  
the  source spectral transitions in the 3$-$200 keV energy range.





\section{Overall pattern of X-ray properties \label{overall evolution}}

\subsection{Hardness-intensity diagram \label{hid}}


To study the properties of 4U~1820-30 during the spectral transitions when the luminosity changes
we   use  hard color (10-50 keV/3-50 keV) (HC) versus  the 3-60 keV  flux  measured in units of $10^{-9}$ erg s$^{-1}$ cm$^{-2}$ [hardeness-intensity diagram (HID)].

In Figure~ 6 in TSF13
we demonstrate flux ratio HC 
versus  the 3-60 keV  flux 
using the {\it RXTE} data. 
As it appears from this Figure, 4U~1820-30 shows a ``J'' like diagonal shape in this diagram 
with upper $short$ and lower $elongated$ branches, which are joined at the lowest flux point. The spectral branches are indicated for the $island$ state (IS), the {\it lower banana} state  (LB) and the {\it upper banana} (UB) state. 
The direction of IS$\to$LB$\to$UB transition is shown by an arrow.
The electron temperature 
$kT_e$  changes 
from 21 keV to about 3 keV along the arrow line direction (compare with Fig. \ref{norm_T_e}).

The identification of hardness-intensity diagram (HID) states are made using  simultaneous timing and
 spectral analysis,  and  we have revisited
the previous similar {\it RXTE} data analysis  made by  
  Bloser et al.  (2000) and  Migliari et al. (2004).

In particular, 
the hard color (HC) drops from 0.25 to 
0.1 while the 3-60 keV flux is quasi constant when the source propagates  from   IS towards the 
LB. 
On the other hand HC rises from 0.12 to 0.35 with a  simultaneous growth of the 3-60 keV flux when 
 the source goes further from   LB towards  UB. 

\section{Evolution of X-ray  spectral properties during spectral state transitions
\label{evolution}}

A number of X-ray spectral transitions of 4U~1820-30 with luminosity variations have been detected 
by {\it RXTE} during 1996 -- 2009 ($R1$ -- $R7$  sets).
We  investigate common spectral$-$timing signatures  which can be found  for these spectral transition 
events.
The source reveals different behaviors  during {\it high-luminosity} and {\it low-luminosity} events.

\subsection{Evolution of X-ray  spectral properties during high-luminosity events \label{evolution_low}}

In Figures \ref{lc_1996}$-$\ref{evolution_lc_3} we show the
results of  our spectral analysis of the {\it RXTE} observations  applying 
the  $wabs*(blackbody+COMPTB+Gaussian)$ model.
On the top panels we present the {\it RXTE}/ASM  count rate and the model fluxes from 3 to 10 keV and from 10 to 50 keV 
(see  {\it blue} and {\it crimson}  points respectively).
 The  TL  electron temperature $kT_e$ as a function of time  is shown in third panel from above. 
The temperature $kT_e$ changes in  the 2.9$-$4 keV interval during  the time period 
MJD 50490$-$MJD 50700 and
only slightly varies around 3 keV during 
MJD 51200$-$MJD 51500 time interval. 
The $COMPTB$ normalization   $N_{COMPTB}$  and  the normalization of low temperature  $blackbody$ component $N_{BB}$  ({\it crimson and  blue} points respectively) 
are presented in the next-to-bottom panel of 
Figs.~\ref{lc_1996}-\ref{evolution_lc_3}.
One can clearly see that  the  $COMPTB$  normalization $N_{COMPTB}$ correlates with  
variations of  the ASM count rate and  the 3 -- 10 keV model flux. 
On the other hand  the blackbody normalization $N_{BB}$ only slightly varies.


\subsection{Evolution of X-ray  spectral properties during the  low-luminosity events \label{evolution_low}}


Since the late April of 2009, 4U~1820-30 showed a  prominent X-ray low-hard state at energies  less than 10 keV 
as  it was observed by the X-ray monitors on  {\it RXTE} and $Swift$.  We display  the characteristics of
the low-hard  state obtained using the  {\it RXTE}  data in  Figure.~\ref{evolution_low}. 
Since 24 April 2009 (MJD 54945), the source was steadily brightening in the 15 -- 50 keV band   of the Swift/Burst Alert Telescope (BAT), with a daily average 
 0.032$\pm$0.002 cts/cm$^2$/s (145 mCrab)~\citep{Krimm09}. 
In contrast, the highest   count
 rate detected by the $Swift$/BAT was  0.14 cts/cm$^2$/s.   

The 
ASM/{\it RXTE} and PCA/{\it RXTE}  light curves showed  that 4U 1820-30 were 
in an extended  low state from MJD 54944 to MJD 54982.
The ASM count rate took a sharp 
drop at  MJD 54944  while the flux began rising in the BAT monitor. 
The ASM count rate 
was very low,
approximately 6.0$\pm$0.5 cts/s during this low state period, with respect to   an usual average count rate
of $\sim$20 cts/s. During the same time period 
the {\it RXTE}/PCA count rate decreases 
from $\sim$4000 cts/s
to $\sim$1000 
cts/s.
This kind of the long-time low  state was not  observed  from  4U~1820-30
over the 10 -- 15 year period. The typical low state duration varies 
from 1 to 2 weeks. 

We also establish  that  the X-ray spectra of this source  over
the $low$ luminosity state (MJD 54955 -- 54982)  are quite stable in terms  of  the $CompTB$ normalization value.
But the electron temperature $kT_e$ of the $Comptonised$ plasma increases from 3 keV up to 20 keV during MJD 54950 -- 54960 period and after that  $kT_e$ gradually decreases again to 3 keV when the  luminosity rises at the end of  the quasi-plateau (at MJD 55000).
In Table 4 we report  the best-fit parameter values. 
During the IS-B 
transition period we find that 
the photon index $\Gamma$ (or the spectral index $\alpha=\Gamma-1$) is almost constant, i.e. only slightly varies  around 2 (see the combined Figure 10 in TSF13).

\subsection{Spectral state transitions in 4U~1820-30}

In 4U~1820-30 
 the hard-soft  state transition is observed 
as 
$kT_e$ decreases from 15 -- 20 keV to 3 keV (see Fig. \ref{evolution_low}).
The outburst hard-to-soft state transitions are seen when 
 a supply of  the  soft photons  flux $N_{COMPTB}$ dramatically increases 
  (see Fig. \ref{norm_T_e} and~ Fig. 6 in TSF13).
 In general, following FT11, 
we
consider the spectral state transitions in terms of  the  $kT_e$ change. 
Thus the hard state is seen when  the electron temperature  reaches maximum, $kT_e^{max}$, whereas   the soft state is observed when  the  electron temperature, $T_e^{min}$ reaches minimum. 
Note that 
$kT_e$ is 
well determined by  the  high energy cut-off  of the spectrum, $E_{cut}$ and can be well established by the spectral fits to the data. This spectral state
definition 
is based on $kT_e$ (or $E_{cut}\sim2kT_e$).
Unlike what occurs in the case of NS binaries, 
for  a black hole case  one can relate a spectral state change to the spectral (photon) index change 
[see \cite{st09}, hereafter ST09].

We test the hypothesis  of  $\Gamma_{appr} \approx 2 $ using  $\chi^2$-statistic criterion.
We calculate the distribution of $\chi^2_{red}(\Gamma_{appr})=\frac{1}{N}\sum_{i=1}^N\left(
\frac{\Gamma_i-\Gamma_{appr}}{\Delta\Gamma_i}
\right)^2$  versus of $\Gamma_{appr}$.
and we  find  a sharp minimum of  function $\chi^2_{red}(\Gamma_{appr})$ at 1 
 when $\Gamma_{appr}=1.99\pm0.01$ 
and $\Gamma_{appr}=1.99\pm0.02$ with   confidence levels of  67\% and 99\% for 234 d.o.f. respectively (see the related Figure of  
$\chi^2_{red}(\Gamma_{appr})$ for 4U 1728-34 in ST11).
The 
index  $\Gamma$  is  almost constant  when $kT_e$ 
(see Fig.
 \ref{gam_Te})  and  the $COMPTB$
normalization,  $L_{39}^{soft}/d^2_{10}$  changes (see below).
 FT11 based on the analysis of {\it Beppo}SAX  data  propose  that  $\Gamma$ is about 2 for quite a few   
NS sources.
FT11 also define  the spectral state using a value 
$kT_e$  and they demonstrate  that $\Gamma=2\pm 0.2$ (or 
$\alpha=1\pm 0.2$)  when  $kT_e$ varies in  the 2.9$-$21 keV interval.

It should be noted that not all NSs shows spectral state transitions but  quite a few NSs exhibit 
them. 
For instance, so called atoll-sources (such as 4U~1820-30), usually demonstrate 
IS -- B transitions.  Particularly during such transitions, it can be possible to differ NS from BH. Specifically, 
NSs and BHs show drastically  different variations of spectral characteristics. 
NSs, as examples of 4U~1820-30 and 4U~1728-34, indicate to variabilities of $kT_e$ and mass accretion rate $\dot M$ 
along with a quasi-constant 
index  $\Gamma$ about 2. Meanwhile, BHs demonstrate a  monotonic growth of  $\Gamma$ when 
$\dot M$ increases and 
succeeded by its final  flattening (saturation) [see ST09]. 

\section{Spectral$-$timing correlations  during spectral 
 state transitions \label{transitions}}

We analyze the {\it RXTE} light curves
applying  the {\it powspec} task taken from
FTOOLS 5.1. We implement the timing analysis {\it RXTE}/PCA data which  we perform in  
the 13-30 keV range  applying the {\it event} mode  
with  time resolution 
of 1.2$\times 10^{-4}$ s. We make 
power density spectra (PDS) in  0.1$-$500 Hz  frequency range
with 0.001$-$second time resolution. The Poissonian statistics contribution
 was subtracted. 
 We apply QDP/PLT 
 package\footnote{http://heasarc.gsfc.nasa.gov/ftools/others/qdp/qdp.html} for PDS modeling.



\subsection{Spectral and timing properties during low luminosity  state transition \label{transitions_low}}

In Figure~12 in TSF13
we present  
a generic behavior of
X-ray timing$-$spectral characteristics for the {\it low} luminosity state at $R7$ (2009) transition events. 
We plot PDSs ({\it left} column) 
along with 
the $EF(E)-$ spectral  diagram ({\it right} column) 
for six moments   at MJD = 54947.6/54956.5, 54958.6/55002.6 and 54997.7/54960.36, covering different 
transition phases.  
At the $bottom$ we demonstrate PDSs for the 15$-$30 keV  energy  range ({\it left} column) and plot along with the $E*F(E)-$spectral diagram ($right$ column)  for A$-$C time events (see upper panel). 
All points [(events) A $red$ (ID 94090-01-01-00), A $blue$ (ID 94090-01-02-03), B $red$ (ID 94090-01-02-02), C $red$ (ID 94090-01-04-03)], 
except  B ($blue$) and C ($red$), are related to IS 
(broadband noise, no VLFN). 

PDSs  denoted by  B ($blue$, ID 94090-02-02-00) and C ($red$, ID 94090-0103-00) 
exhibit the {\it island$-$lower banana} state transition. 
For  the $blue$ PDS the VLFN  (very low frequency noise) appears as a band-limited 
noise component which   transforms into  QPO (broad Lorentzian with $\nu_h$ centroid frequency at 7 -- 10 Hz).
We present power density spectra (PDSs)  
(panels A1, B1, C1) along with 
the   corresponding $E*F(E)-$spectral diagrams (panels A2, B2, C2). 
The  related PDSs and energy spectral data  are shown by blue and red  points respectively.
On the left panel we  also show the electron temperature $kT_e$ associated with a given PDS.

\subsection{Spectral and timing properties during high luminosity  state transition \label{transitions_high}}


To compare with  Figure~12 in TSF13,
we show an evolution of spectral$-$timing 
characteristics
during the {\it high} 
luminosity  state  in  Figure.~\ref{ev_PDS_SP_high}. 
Here on the $top$ panel we display the ASM light curve 
during $high$ luminosity interval at $R3$ (1999) transition events. Red/blue points 
A, B, and C are related to the moments at MJD = 51283.6/51300, 51313.7/51330.5 and 51389.4/51396.26, covering different 
transition phases.  
On the  $bottom$ {\it left} and {\it right} panels we present PDSs for 15-30 keV  energy range 
and the $E*F(E)-$spectral diagram correspondingly,
for A ($red$, top), B ($blue$, middle) and C ($blue$, bottom) points  of X-ray light curve
(see the upper panel).   
All points are related to the  $banana$ state with relatively strong broadband noise and VLFN with QPOs which is at  $\nu_l\sim$ 6 -- 7 Hz  for C moment ($red$). 
The PDSs in panel C ($blue$) and ($red$) illustrate  the {\it island}$-${\it lower banana} state transition. Here the VLFN appears, band-limited noise component transforms into  QPO [a broad Lorentzian with $\nu_h$ centroid frequency at 7$-$10 Hz, C $red$ (40017-01-12-00)].
The power spectra (panels A1, B1, C1) correspond to $E*F(E)$ diagrams (panels A2, B2, C2).  
The corresponding  energy spectra of 4U~1820-30 are related to the electron temperature of 3 keV.

In Figure~\ref{PDS} we illustrate a  typical power spectrum of 4U~1820-30 for different  X-ray spectral  states (shown on the right panel). The electron temperature  values of corresponding energy spectra are indicated at the right 
vertical axis. The power spectra in the extreme island state (EIS), island state (IS, multiplied by factor $10^{-2}$ for clarity),
lower left banana state (LLB, $\times 10^{-4}$), lower banana state (LB, $\times 10^{-6}$) and upper banana state (UB, $\times 10^{-8}$) 
are presented from the {top} to the {\it bottom}. The histograms show the best fits to  the power spectra, which  consist of 
three components: VLFN 
the peaked 
noise component,  
low-frequency  QPOs 
($\nu_{l}$ and $\nu_{h}$) and high frequency feature $\nu_{hHz}$
(see van Straaten, van der Klis \& Mendes 2003 for details of terminology). 

\subsection{Comparison of spectral and timing  characteristics of  {\it atoll} sources 4U~1820-30, GX~3+1 and 4U~1728-34} 

In this Paper, we also  study the correlations of X-ray spectral$-$timing characteristics and $\dot M$
in a number of  $atolls$ 
during their spectral transitions searching 
for similarities and differences between 
$atoll$ 
sources.  In this way we can present a comparative analysis for three {\it atoll} sources: 
4U~1820-30, GX~3+1 and  4U~1728-34 applying  the same spectral 
model which consists of low temperature $Blackbody$, $Comptonized$ continuum and $Gaussian$ line component.

\subsubsection{Constancy of the photon index}

We demonstrate that {\it atolls}  4U~1820-30, GX~3+1,  4U~1728-34 show  a similar pattern of  the photon index  $\Gamma$ vs  $\dot M$
(or $N_{COMPTB}$).
 Namely, 
 the photon  index $\Gamma$ only slightly varies   
around 2 (see Fig.~\ref{gam_norm_3obj}). 
Following to FT11, ST11 and ST12,  we can suggest that 
the cooling flow of  soft disk 
 photons is much less less than  
the energy release in the transition layer (TL)  for each of these
three sources.



\subsubsection{The difference and similarity of  $kT_e$ ranges in 4U~1820-30,  GX~3+1 and 4U~1728-34 }

\begin{figure}[ptbptbptb]
\includegraphics[scale=1.2,angle=0]{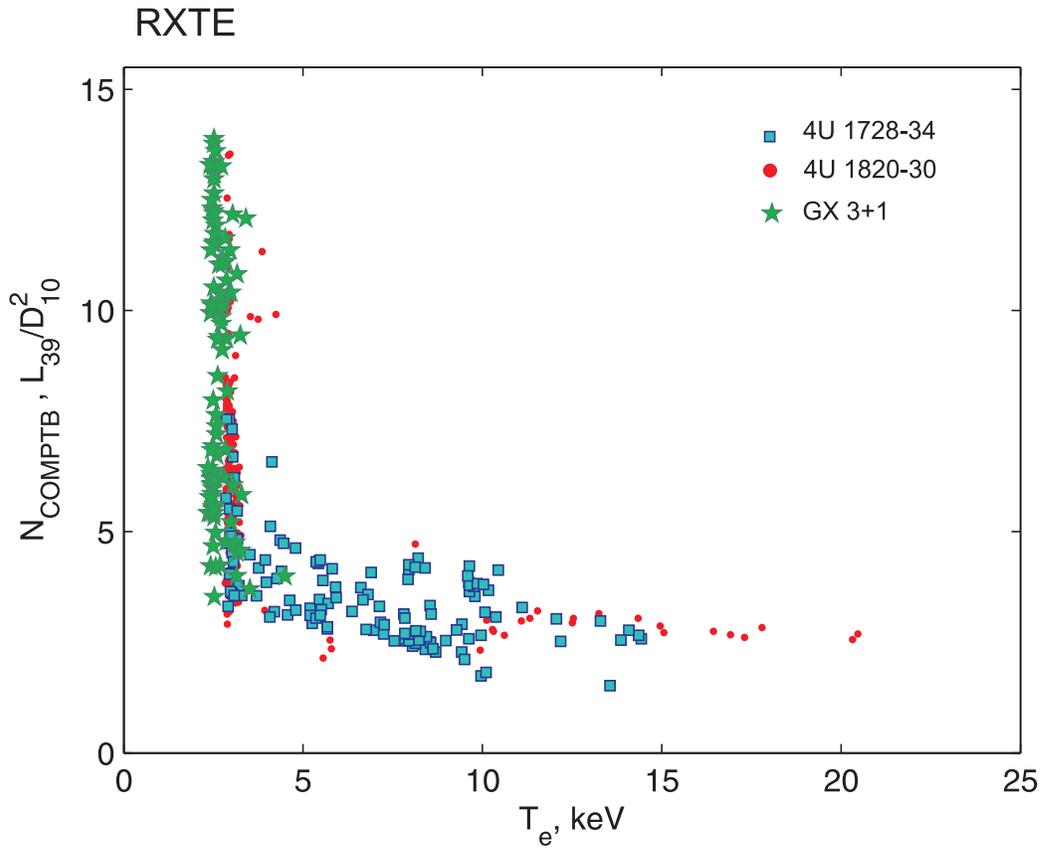}
\caption{Comptb normalization measured in units of $L^{soft}_{39}/D^2_{10}$  
versus the electron temperature $kT_e$ (in keV) obtained using the best-fit  spectral model 
$wabs*(blackbody+Comptb+Gaussian)$ for  {\it atoll}  sources 4U~1820-30 ($red$), 
GX~3+1 ($green$, taken from ST12) and 
4U~1728-34 ($blue$, taken from ST11)  for  {\it RXTE} data. 
} 
\label{norm_T_e}
\end{figure}
\begin{figure}[ptbptbptb]
\includegraphics[scale=0.8,angle=0]{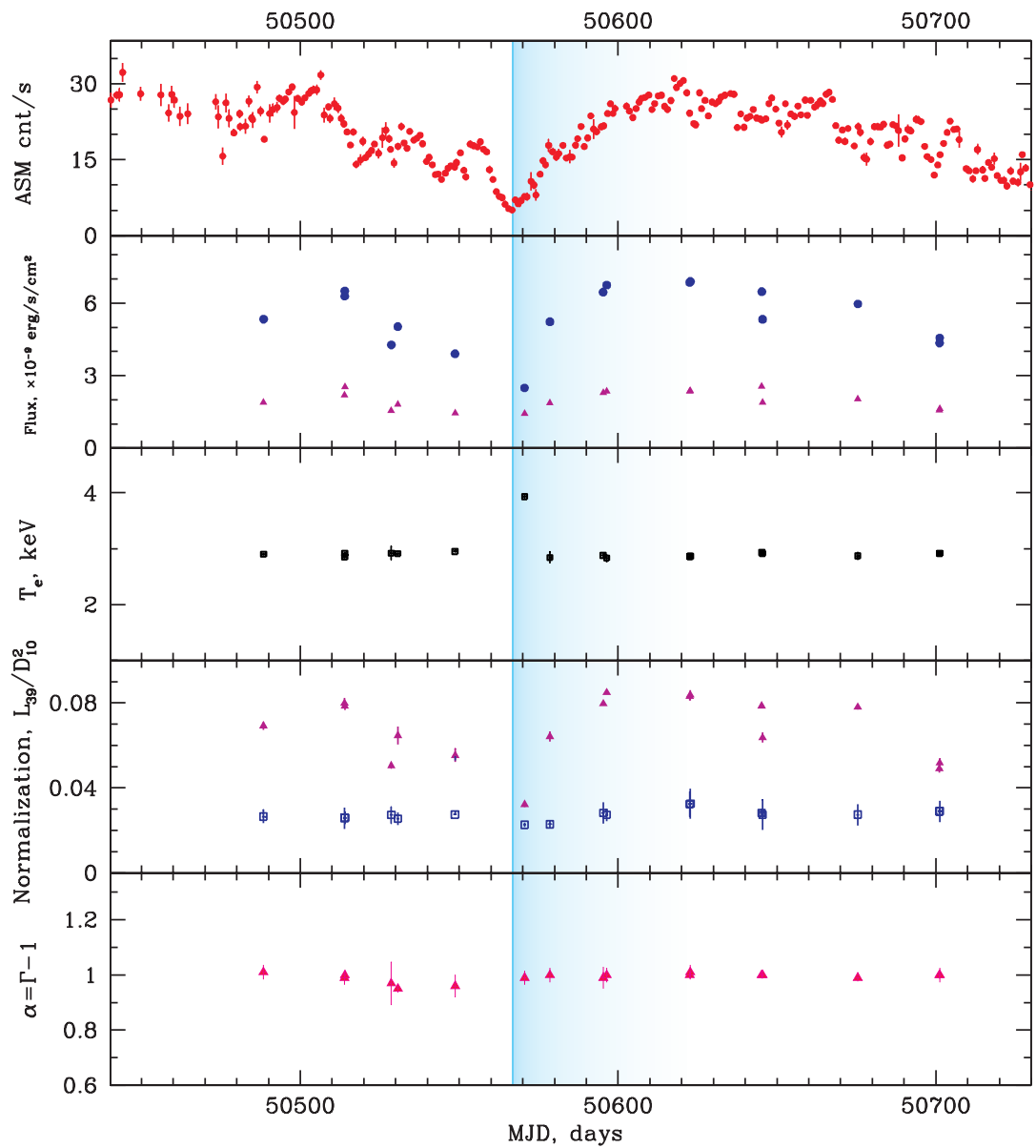}
\caption{
{\it From Top to Bottom:}
Evolutions of the {\it RXTE}/ASM count rate, the model flux in 3-10 keV  and 10-50 keV energy ranges 
({\it blue and crimson} points respectively), the electron temperature $kT_e$ in keV, 
 and $Comptb$ and $blackbody$ 
normalizations ({\it crimson} and $blue$ 
respectively)  during 1996 -- 1997 
transition set ({\it R1 -- R2}). 
 The rising phases of the $local$ ($mild$) transitions are marked with blue vertical strips. 
}
\label{lc_1996}
\end{figure}
\begin{figure}[ptbptbptb]
\includegraphics[scale=0.8,angle=0]{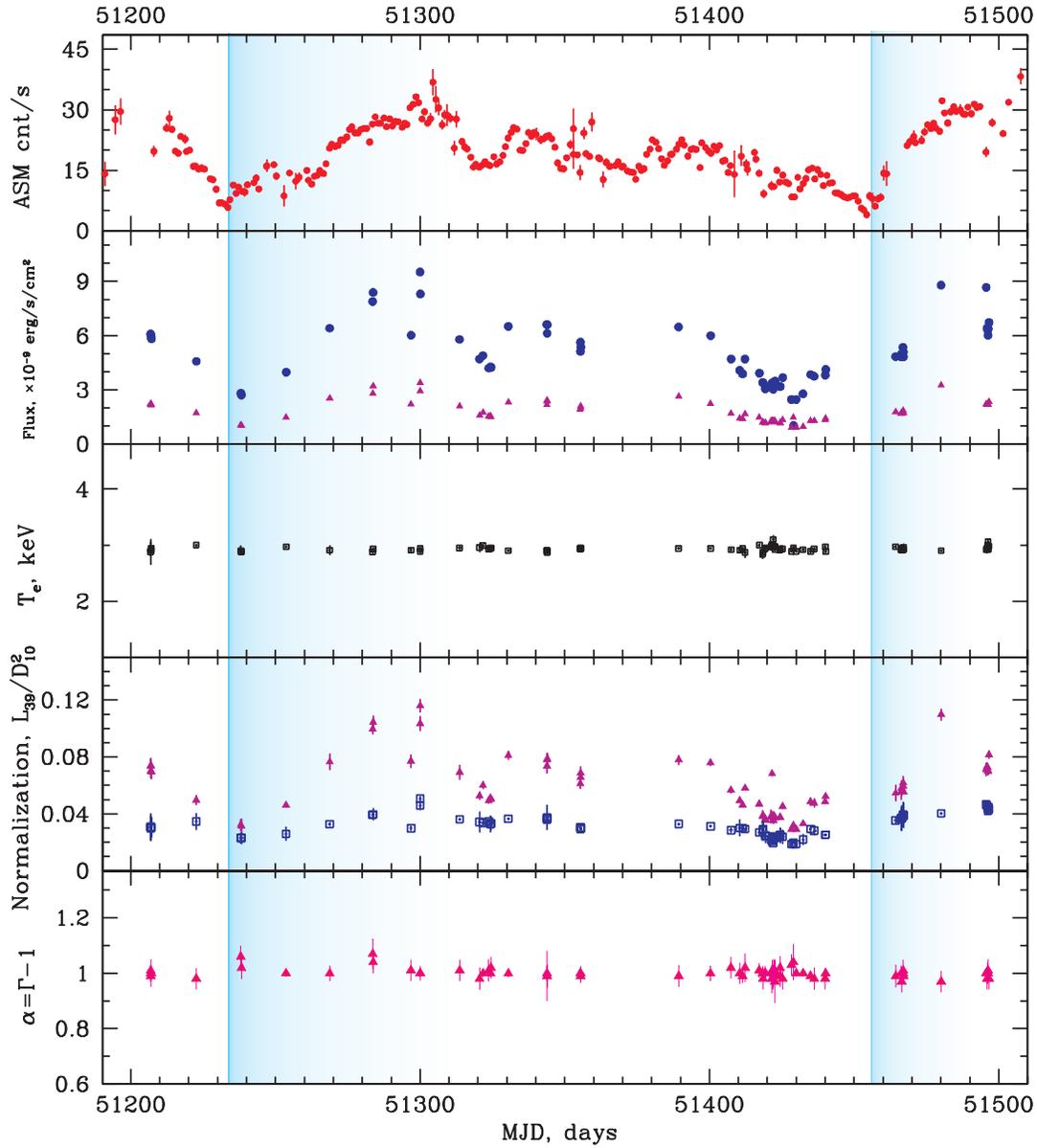}
\caption{ Similar to  that presented in Fig. \ref{lc_1996} but  for the {\it RXTE} 1999 
transition set 
  {\it R3}.
}
\label{evolution_lc_3}
\end{figure}

One can see from  Figure~\ref{T_e_vs_f_comp},   that the ranges of 
$kT_e$ for an individual state evolution of these three sources are  different. 
The electron temperature $kT_e$  
  changes in 4U~1728-34
  from 3  to 15 keV whereas  $kT_e$ 
 varies within much narrow range of $kT_e$ around 3 keV in  GX~3+1. 
In turn, source 4U~1820-30 demonstrates a wider interval of $kT_e$ in which $kT_e$ varies  from 2.9 keV to 21 keV
similar to some extent to the temperature change  in 4U~1728-34.  
Note, that in a low temperature regime  
4U~1820-30 and GX~3+1 are similar in terms of  normalization $N_{\rm comptb}=$(4 -- 15)$\times L_{39}/D^2_{10}$, or mass accretion rate, (see  Fig. \ref{norm_T_e}) 
and Comptonized fraction $f=$ 0.2 -- 0.8 (see Fig.~\ref{T_e_vs_f_comp}).
While  
4U~1820-30 and  4U~1728-34  are similar for a range of the normalization  
$N_{\rm comptb}=$(2 -- 4)$\times L_{39}/D^2_{10}$ (see Fig.\ref{norm_T_e}) 
and  $f=0.5-0.8$ (see Fig.~\ref{T_e_vs_f_comp}) for high temperature regime.
However, in contrast to  4U~1728-34, the source  4U~1820-30 has an additional branch  of  intermediate temperatures 
(8 -- 12 keV) when the Comptonized fraction is relatively low, $f<0.5$
( see Fig.~\ref{T_e_vs_f_comp}). 
Note that all objects have a common temperature interval 3 -- 4 keV when $N_{\rm comptb}$ (or mass accretion rate) is relatively high. 

Thus according to FT11, ST11, ST12 and the present study, the electron
temperature $kT_e$ for  {\it atolls} and $\it Z-$sources varies  in the  2.5$-$25 keV range. 
Specifically, the change of $kT_e$  around 3 keV is similar   for all three source 4U~1820-30, 
GX~3+1 and 4U~1728-34.
The minimum  value of $kT_e$   at 2.5 keV occurs 
at the  peak luminosity for 4U~1728-34 (see ST11),  during a local rise of luminosity for GX~3+1 (ST12) 
and  at high luminosity phases 
for 4U~1820-30.

For all of these  three objects, the values  of  color  seed photon temperatures 
$kT_s=1.1-1.7$ keV  and blackbody temperatures $kT_{BB}\simeq$0.6 keV are 
  comparable  (see Table 5).
Contrary,  the variability extent of 
 $kT_e$  is not similar.
The reason for that  difference of  electron temperature ranges is evident.
Sources 4U~1820-30 and 4U~1728-34 show a complete cycle of state evolution: 
 {\it island}$-${\it lower banana} (LB)- {\it upper banana} (UB) stages for 4U~1820-30 
and   for 4U~1728-34 it is  {\it extreme island} state (EIS)$-${\it upper banana} (UB) state   (see Di Salvo et al. 2001; ST11).
But GX~3+1 demonstrates  a short evolution behavior on the CCD from LB to UB. 
This evolution picture is also clear  from  Figure~\ref{T_e_vs_f_comp} which  shows that  the track of  GX~3+1 is only  a part of the full track
[see definition of a state sequence  and the standard $atoll$-{\it Z} scheme in Hasinger \& van der Klis (1989)].

Note, that  
4U~1820-30 shows almost the same  $kT_e$ range as that in 4U~1728-34 and
almost similar timing evolution. But  clear differences between these 
{\it atolls} 
 one can   see from Figure~17 in TSF13
 where we  show 
spectral hardness (10 -- 50 keV/3 -- 50 keV) vs flux in  3 -- 60 keV range. 
In fact, 
4U~1728-34 ({\it blue} points) is  fainter and harder  and demonstrates much wider spectral hardness range than  that in
 4U~1820-30 ({\it red} points).

 
\subsubsection{Comparison of spectral evolution as a function of the
$luminosity$
for 
4U~1820-30, GX~3+1 and 4U~1728-34}

Now we present  a comparison  
  of X-ray spectrum  evolution for 
sources 4U~1820-30, GX~3+1 and
 4U~1728-34 based on $luminosity$ value which is presumably proportional to $Comptb$ normalization and, consequently, to mass accretion rate 
taking into account that their distances to the Earth 
are similar
(see Table 5). 
For 
4U~1820-30 the distance range is within of  5.8 -- 8 kpc (Shaposhnikov \& Titarchuk, 2004), whereas 
for 4U~1728-34 and GX~3+1  the distances are estimated as 4.5 kpc~ and 
4.2$-$6.4 kpc~respectively [see \cite{par78} and  \cite{kk00}].   

We show 
the $CompTB$ normalization  (related to the soft photon $luminosity$ value)
for these 
sources as a function of 
$kT_e$ in Figure \ref{norm_T_e}. 
4U~1820-30 subtends a wider interval in 
$CompTB$ normalization than that for  
 4U~1728-34.
Note, that in the high luminosity state (or $N_{\rm comptb}$) 
 4U~1820-30 is similar to GX~3+1:  
$N_{\rm comptb}=$ (4 -- 15)$\times L_{39}/D^2_{10}$,  
  Comptonized fraction $f=$0.2 -- 0.8 (see  Fig.~\ref{T_e_vs_f_comp}) and  the electron temperatures $kT_e$ are 
low variable around  3 keV.
While in the low luminosity state  4U~1820-30 is closer to source 4U~1728-34: 
$N_{\rm comptb}=$ 
(2 -- 5)$\times L_{39}/D^2_{10}$ (see Fig.\ref{norm_T_e}),   $f=$0.5 -- 0.8 and  
$kT_e$ changes from 5 to 20 keV (see Fig.~\ref{T_e_vs_f_comp}).

\subsubsection{The difference and similarity of time scales of state evolution for 4U~1820-30, GX~3+1 and 4U~1728-34}

We should  point out  that all these three $atoll$ source show 
the transitions between low luminosity and high luminosity states over
different time scales. 
Specifically, the time scales of X-ray flux variability for 4U~1728-34, 4U 1820-30 and GX~3+1 probably 
dictated by variability of mass accretion rate, are $\sim$10 d, 100 d and 1000 d, respectively. 
However,  these  sources  demonstrate  LB -- UB transition  and make it in the narrow 
interval of low temperature $kT_e$ (around 3 keV) and  during  the same short  time interval  (hours -- day). 
We remind the reader that the comparison between these three sources  is facilitated  by the fact  that they show almost  
the same $kT_{BB}$ and $kT_{s}$ temperature values and they are located  at approximately the same distance. 
The only difference of spectral evolution of  these objects is related to  different ranges  of the electron temperature  of the  Comptonized component.  


\subsubsection{Correlation of illumination parameter  $f$ versus electron temperature $kT_e$ and its relation with different stages in the color-color diagram}


Using Table 5 one can see that 
the ranges of the best-fit illumination  fraction $f$ are   $0.2-1.0$, $0.2-0.9$ and $0.5-1$ for 4U~1820-30,  GX~3+1 and  4U 1728-34 correspondingly. 
These values of $f$   indicate to different geometry  of the transition layer (TL) and thus to different illumination  
for these X-ray sources. 
In  Figure~\ref{T_e_vs_f_comp}  we demonstrate that the electron temperature $kT_e$  directly correlates   with  a sequence  of CCD states, EIS-IS-LLB-LB-UB  [see \cite{hasinger89} for this CCD classification]. 
Note that ST12 reveal a 
relation 
between 
spectral states,  $kT_e$ and $f$ for $atoll$ sources 
GX~3+1 and 4U~1728-34.  We show these $kT_e-f$  relations  for these two atolls  in  Figure~\ref{T_e_vs_f_comp}.
The direction in which the inferred $\dot M$  increases is indicated by arrows.

Now we present  three different tracks 
on the  $kT_e-f$  diagram
for 
 three source 4U~1820-30, GX~3+1 and 4U~1728-34 and show how these tracks are related to
the standard  CCD sequence (see  Fig.~\ref{T_e_vs_f_comp}).  
The track of 4U~1820-30 consists of three segments (branches) related to $kT_e$: high (12 -- 21 keV), intermediate (7 -- 12 keV) and 
low temperature (2.9 -- 6 keV) ones, wherein each segment  has a negative correlation of $kT_e$ and  $f$.
In turn, GX~3+1 demonstrates the only, so called, {\it low temperature}  branch track. Namely, when the fraction 
$f$ increases,  $kT_e$ decreases 
from 
 $\sim$ 4.5 keV to $\sim$~2.3 keV.
 While for  4U~1728-34 we see a more complicated 
 pattern, but 
in contrast to 4U~1820-30 and GX~3+1, has a segment with positive correlation of $kT_e$ vs $f$ from 4 to 12 keV. 
Specifically, at the high temperature state (EIS), 
 $f$ only slightly changes from 0.9 to 1 when  $kT_e$ decreases. 
 As  $kT_e$ further drops  from
12 keV to 4 keV, $f$ also drops from 0.9 to 0.5. 
Finally, 
 $f$ goes up  from 0.5 to 0.8  when the source enters to  the low-temperature state (LB-UB). 
As a result, we demonstrate  that the CCD state evolution 
can be also seen using the $kT_e-f$ correlation. 

\subsection{Comparison of spectral hardness diagrams for atolls 4U~1820-30, GX~3+1 and 4U~1728-34 \label{ccd}}

We use the plot HC (10-50 keV/3-50 keV) versus the 3 -- 60 keV  flux   
in form of 
HIDs for three sources:
4U~1820-30 ($red$), GX~3+1 ($green$) and 4U 1728-34 ($blue$)  (see Fig. 17 in TSF13)
to compare   transition properties of these {\it atolls}  in terms of their flux (or luminosity).
In fact,
4U~1820-30 shows a ``J'' like inclined (or diagonal) shape in HID with upper $short$ and 
lower $elongated$ branches (see  Fig. 6 in TSF13).
The $short$ branch is close to the low luminosity state, whereas the $elongated$ branch covers
the wide luminosity range. 

Our comparative analysis 
of HID track  branches for   4U~1820-30, GX~3+1 and 4U~1728-34
 indicates to similar physical properties of these objects. The spectral and timing characteristics are 
very similar along corresponding segments. Specifically, the $short$ branch of a ``J'' like track 
of 4U~1820-30 is adjacent to low luminosity  area of 4U~1728-34 ($blue$ points)  and it is related to
a high electron temperature   regime 
of  4U~1820-30, as in 4U~1728-34. 
In turn, the $elongated$ branch of 4U~1820-30 is closer to GX~3+1 ($green$ points) branch area and it is associated 
with low electron temperatures $kT_e$ (3 -- 4 keV) and softer spectra, which  are also seen  in GX~3+1.

Note that among considered {\it atolls}  superbursts are only  observed 
 in GX~3+1 and 4U~1820-30 during $elongated$ branch. Furthermore, superbursts are 
detected at low$-$soft states, i.e. during low luminosity interval of light curve when electron 
temperature $kT_e$  is about 4 keV.  Thus 4U~1820-30 shows a  property similar to GX~3+1 and  it is situated at
an {\it intermediate position} between 4U~1728-34 and GX~3+1 
in terms  its  luminosity.  This observational  fact can  be related to the same   {\it intermediate} rate of mass transfer in these two sources, 4U~1820-30 and GX 3+1 [see also  a review by van der Klis (1994)].
The comparison of HIDs allows to diagnose physical properties of different objects with adjacent HID tracks.

 



\section{Discussion \label{disc}}

\subsection{Stability of photon index 
is a signature of NS source \label{constancy}}

Thus we  demonstrate that  the photon spectral  index only slightly varies around 2 
  using   numerous  observations of NS sources  4U~1820-30, 
GX~3+1 (ST12)  and 4U~1728-34 (ST11) by   {\it Beppo}SAX and {\it RXTE}.  
In Figures \ref{gam_Te},  
\ref{gam_norm_3obj} and \ref{bh_ns_examples} ($left$ column) we show   
$\Gamma$ as a function of  the spectral  
model parameters: 
$kT_e$ (in keV),   $N_{COMPTB}-$normalization, 
and illumination fraction $f$.  
These results for NS 4U~1820-30 have been obtained when  we apply our  thermal Comptonization 
model to {\it Beppo}SAX and extensive 
{\it RXTE}  observations. 
FT11 and ST11 also investigate  and find 
the photon (energy)  index  stability  
in other observations of NS binaries. 
 We   explain this index stability 
 using  the Comptonization  model. 
Namely, the photon (energy) index is almost constant  
when the soft photon flux illuminated  the transition layer (TL)
is much less than the  gravitational energy release in TL (see e.g. ST12).
This model  of the index stability  can  probably resolve    the  index stability effect  now clearly established 
in these three NS sources  using  extensive {\it Beppo}SAX and 
{\it RXTE} observations.

\subsection{On the {\it hard tail} origin in {\it atoll} source 4U~1820-30 \label{constancy}}

The radio emission detected from 4U 1820--30 \citep{Migliari04} 
suggests the presence of a jet, which may also generate an  extended power-law X-ray emmision. 
In this case,  the power law can be a result of the inverse Compton effect on 
 nonthermal electrons of the jet. 
Note, X-ray nonthermal power-law tails  are also observed  in soft states of BHs 
see for example  a review 
by  ST09;
see also \cite{mcconel} and \cite{wardz} on the detection of the extended hard tails in the hard states of BHs,  
Cyg X-1 and GX 339-4 respectively, 
and NS Z-sources [see
\cite{Di_Salvo2000}; \cite{Far05}; \cite{amico}; 
\cite{asai}]. 
However  these extended hard tails  are also  found   in {\it atolls}  [see e.g.  \cite{Piraino99}].

 Additive models  that have been applied to fit the spectra  of 4U 1820-30, need to use an additional power-law component (pure one or as a component of CompPS) to describe a hard spectral tail above 
80 keV [see e.g.  \cite{tarana06}].
However  such an approach invokes an unknown  non-thermal origin of {\it hard tail} emission. On the other hand our suggestion allows  us to explain   X-ray spectra of 4U1820-30 in {\it all spectral states} using    the same model without  a specific composition of the model components at different states. In fact, in our model  (see \S 5.2 and \S 6.1) we describe the {\it hard tail} emission using the thermal Comptonization component in which the TL electron temperature $kT_e$  increases up to 20 keV 
 and the illumination factor $f$ decreases as the source goes to the hard state (see Figs. \ref{sp_compar}, \ref{T_e_vs_f_comp}).

\section{Conclusions \label{summary}} 

We analyze  the X-ray  spectral and timing characteristics of  4U~1820-30 observed 
during the hard-soft  state transitions.
We find  a number of  spectral transitions in 4U~1820-30  using  {\it BeppoSAX}  and 
{\it RXTE} data.  

For our  investigation  we take an advantage  
 the $Beppo$SAX broad  spectral extension 
over the 0.3$-$200 keV  range and   abundant    $RXTE$ observations taken  
in the 3$-$200 keV energy coverage.

We demonstrate  that the X-ray broad-band spectra  can be successfully  fit by   composition of
 the  {\it Blackbody},  Comptonization  ({\it Comptb}) and 
 {\it Gaussian}$-$line  components for  {\it all spectral states}. Also we  show  an observable 
 relation of the photon index $\Gamma$ and  the  normalization of the Comptonized component,  {\it Comptb} which is proportional to $\dot M$.
We demonstrate  the stability of 
the photon index 
$\Gamma\sim 2$  when  the source goes from the hard state to 
soft state, in other words  when the electron temperature  of Comptonized region (TL), $kT_e$  decreases  from 21 to 3 keV  (see 
Fig.~\ref{gam_Te}). 

We  also show  that $\Gamma$  only slightly varies with  the {\it Comptb} normalization 
($\propto \dot M$).
Note,  this stability of the index in NS sources   has been recently suggested   for   a  
 number of other  NSs,  Sco X-1, Cyg X-2,  GX 17+2, GX 3+1,
GX 340+0,  GX 349+2, X 1658-298, 1E 1724-3045, GS 1826-238,  which were  analyzed 
 using  {\it Beppo}SAX data 
(see details in  FT11, ST11, ST12).
The use of the disk  {\it seed} photon normalization, ({\it Comptb}),  which is proportional to $\dot M$,
is fundamental in order to find
the stability of $\Gamma$ 
 during  the hard-soft state transition. We do find the stability (constancy) 
of the photon index of Comptonized component versus both the {\it Comptb} normalization and the electron 
temperature $kT_e$ about  2 for  all spectral  states.  In our analysis of NS sources (see FT11, ST11, ST12, Seifina et al. 
2013 and this Paper) we do not find any particular case in which the photon  index $\Gamma$ changes beyond the limits $2\pm 0.1$.   Thus this index stability  
can be  taken as an intrinsic property of  neutron star (NS) binaries ({\it as a NS signature}),  which  is drastically  
different   from that in black hole binaries [e.g., GX~339-4, GRO~J1655-40, XTE~J1650-500,  XTE~J1550-564, 4U~1543-47, XTE~J1859+226,
H~1743-322, (ST09), GRS~1915+105 (TS09), SS~433 (ST10)], where $\Gamma$ monotonically rises  during  the hard$-$soft state transition and it follows by its saturation
at high $\dot M-$values 
(see Fig.~\ref{bh_ns_examples}). 
In Figure~\ref{bh_ns_examples} we show the $\Gamma-\dot M$ correlation 
for a number of BHs 
($right$ column) and that $\Gamma$ is almost independent of $\dot M$ 
in NSs ($left$ column). 
 Indices $\Gamma$  in BHs
 show clear correlation with
$\dot M$, or with  the  normalization $L_{39}/D^2_{10}$ [where $L_{39}$ is a flux of soft (seed) photons].
 The $\Gamma-\dot M$ correlation is followed by $\Gamma-$saturation 
 when mass accretion rate $\dot M$ 
exceeds the Eddington limit. 
The behavior of $\Gamma$ vs $\dot M$ for a considered sample of  NSs (4U~1820-30, 4U~1728-34 
and GX~3+1) is drastically different from that for  given examples of BHs.

A relatively  wide interval of  the illumination fraction $f=0.2-1$ which we obtain in the framework of our 
model,   
point to  
  variable 
 soft (disk)  photon illumination of the transition layer   in  4U~1820-30.
Using {\it Beppo}SAX  data we also find  
two types  of blackbody photons. One type is characterized by color temperature of 0.7 keV, which is typical for the disk photons and another one is related to 1.3 keV,  which can be associated  with NS surface temperature. 

We detect an evolution of 6 -- 20 Hz QPOs and noise components 
during the $island$ -- $banana$ 
state evolution (LLB-UB)  (see Fig. \ref{ev_PDS_SP_high}).


{\it Our observational results establishing the constancy of the photon index  $\Gamma$ in 4U 1820-30 confirm  the  theoretical arguments of FT11 and ST11  that  the TL energy release 
$Q_{cor}$ dominates  
the soft  photon flux illuminating the transition layer (TL)  which  comes from the accretion disk,  $Q_{disk}$.  
 We argue that the found  stability of $\Gamma$   is an intrinsic NS signature
 as in BH binaries  $\Gamma$ monotonically   increases with $\dot M$ followed by its saturation at high values of $\dot M$
(see ST09)}.

LT acknowledges discussion with Chris Shrader and his thorough editing of the  manuscript and also we appreciate comments of the referee which substantially improve a quality of the presented material.

\newpage

\begin{deluxetable}{cccc}
\tablewidth{0in}
\tabletypesize{\scriptsize}
    \tablecaption{The list of {\it Beppo}SAX observations of 4U~1820-30 used in analysis.}
    \renewcommand{\arraystretch}{1.2}
\tablehead{
Obs. ID& Start time (UT)  & End time (UT) &MJD interval}
\startdata
20105004& 1997 Oct. 2 06:45:23 & 1997 Oct. 2 19:09:36 &50723.2-50723.8 \\
20537004& 1998 Apr. 17 04:31:23 & 1998 Apr. 18 02:55:28 &50920.1-50921.1$^1$\\
20537005& 1998 Sept. 23 12:44:56 & 1998 Sept. 24 15:30:05 &51079.5-51080.6$^1$\\
      \enddata
   \label{tab:table}
Reference. 
(1) \cite{kaaret99} 
\end{deluxetable}

\newpage
\begin{deluxetable}{lllll}
\tablewidth{0in}
\tabletypesize{\scriptsize}
    \tablecaption{The list of groups of {\it RXTE} observations of 4U~1820-30}
    \renewcommand{\arraystretch}{1.2}
\tablehead{Number of set  & Dates, MJD & RXTE Proposal ID&  Dates UT &  Ref.}
 \startdata
R1  &    50204-50207  & 10074        & May 1 -- 4, 1996      & 1     \\
   &    50151, 50235 & 10076        & March 9, June 1, 1996 & 1     \\
    &    50371-50386  & 10075        & Oct. 15 -- 30, 1996   & 1, 2, 3, 4    \\
R2  &    50488-50701  & 20075        & Feb. 9 -- Sept. 10, 1997 & 1, 3, 4, 5, 6, 7 \\
R3  &    50920-51467  & 30053, 30057 & Apr. 4, 1998 -- Oct. 16, 1999 & 3\\
  &    51206-51956  & 40017, 40019 & Jan. 28, 1999 -- June 4, 2003 &    \\
R4  &    51996-51999  & 60030        & March 28 -- 31, 2001            & \\
  &    52355, 52429  & 40017        & March 22, 2002; June 6, 2003    &  \\
\enddata
    \label{tab:par_bbody}
References:
(1) \cite{Chou01}; 
(2) \cite{smale97}; 
(3) \cite{kaaret99};
(4) \cite{Zhang98};
(5) \cite{Bloser00};
(6) Ku$\acute s$mierek et al. 2011; 
(7) Shaposhnikov \& Titarchuk (2004);
(8) Migliari et al. (2004);
(9) Krimm et al. (2009) 
\end{deluxetable}

\newpage
\bigskip
\begin{deluxetable}{cccccccccccccc}
\rotate
\tablewidth{0in}
\tabletypesize{\scriptsize}
    \tablecaption{Best-fit parameters of spectral analysis of {\it Beppo}SAX 
observations of 4U~1820-30 in 0.3-200~keV energy range$^{\dagger}$.
Parameter errors correspond to 90\% confidence level.}
    \renewcommand{\arraystretch}{1.2}
 \tablehead
{Observational & MJD, & $T_{BB}$, & $N_{BB}^{\dagger\dagger}$ &$T_s$, & $\alpha=$  & $T_e,$ & $\log(A)$ & N$_{COMPTB}$ &  E$_{line}$,&   $N_{line}^{\dagger\dagger}$ &  $\chi^2_{red}$ (d.o.f.)\\
ID             & day  & keV         &                           & keV  &$\Gamma-1$  & keV    &                                  &                                   &   keV       &                               &                        }
 \startdata
20105004   &        50723.28 & 0.63(3) & 2.33(2) & 1.35(2) & 1.00(4) & 3.25(2) & -0.11(5) & 3.45(2) & 6.7(1) &  0.45(2) & 1.11(364)\\
20537004   &        50920.19 & 0.58(1) & 2.74(3) & 1.24(5) & 1.01(5) & 3.42(2) & -0.10(6) & 4.42(6) & 6.6(5) &  1.38(7) & 0.89(364)\\
20537003   &        51079.53 & 0.69(2) & 2.68(3) & 1.27(4) & 0.99(8) & 3.37(1) & -0.12(4) & 4.65(1) & 6.5(2) &  0.41(5) & 1.2(234)\\
\enddata
    \label{tab:fit_table}
$^\dagger$ The spectral model is  $wabs*(blackbody + Comptb + Gaussian)*smedge$,
 where $N_H$ is fixed at a value 3.00$\times 10^{21}$ cm$^{-2}$ (Bloser et al., 2000); 
$^{\dagger\dagger}$ normalization parameters of $blackbody$ and $Comptb$ components are in units of 
$10^{-2}\times L_{39}/d^2_{10}$ $erg/s/kpc^2$, where $L_{39}$ is the source luminosity in units of 10$^{39}$ erg/s, 
$d^2_{10}$ is the distance to the source in units of 10 kpc 
and $Gaussian$ component is in units of $10^{-2}\times total~~photons$ $cm^{-2}s^{-1}$ in line, 
wherein $\sigma_{line}$ of $Gaussian$ component is fixed to a value 0.8 keV 
(see comments in the text); smeared edge included at 7.7 keV.
\end{deluxetable}

\newpage
\bigskip
\begin{deluxetable}{llllllllllllll}
\rotate
\tablewidth{0in}
\tabletypesize{\scriptsize}
    \tablecaption{Best-fit parameters of spectral analysis of PCA\&HEXTE/{\it RXTE} 
observations of 4U~1820-30 in 3-200~keV energy range$^{\dagger}$.
Parameter errors correspond to 90\% confidence level.}
    \renewcommand{\arraystretch}{1.2}
 \tablehead
{Observational & MJD, & $\alpha=$  & $T_e,$ & $\log(A)$$^{\dagger\dagger}$ & N$_{COMPTB}^{\dagger\dagger\dagger}$ & $T_s$,& $N_{Bbody}^{\dagger\dagger\dagger}$ & E$_{line}$,&   $N_{line}^{\dagger\dagger\dagger}$ &  $\chi^2_{red}$ (d.o.f.)& F$_1$/F$_2^{\dagger\dagger\dagger\dagger}$ \\
ID             & day  & $\Gamma-1$ & keV    &                           &                                       & keV&                                     &  keV       &                                      &                         &                                                & }
 \startdata 
%
%
10076-01-01-00 & 50151.937 & 1.00(2) & 2.87(2) & 2.00$^{\dagger\dagger}$ & 7.13(1) & 1.30(5) &  3.42(1) & 6.27(8) & 1.1(1)  & 1.54(78) & 6.13/2.24 \\
10074-01-01-00 & 50204.375 & 1.0(1)  & 2.93(4) & 2.00$^{\dagger\dagger}$ & 5.5(2)  & 1.0(1)  &  2.31(3) & 6.52(6) & 0.15(3) & 0.92(78) & 4.58/1.61 \\
10074-01-01-01 & 50204.511 & 0.99(2) & 2.91(1) & 2.00$^{\dagger\dagger}$ & 5.3(1)  & 1.08(2) &  2.57(4) & 6.5(1)  & 0.23(3) & 0.93(78) & 4.52/1.58 \\
10074-01-01-02 & 50204.785 & 1.00(2) & 2.95(2) & 0.92(8) & 5.25(9) & 1.05(3) &  2.48(6) & 6.40(4) & 0.14(2) & 1.06(78) & 4.33/1.53 \\
10074-01-02-02 & 50207.381 & 1.00(5) & 2.84(2) & 0.9(1)  & 7.5(1)  & 1.06(5) &  2.15(4) & 6.58(4) & 0.23(3) & 1.09(78) & 5.90/2.06 \\
\enddata
    \label{tab:fit_table}
$^\dagger$ The spectral model is  $wabs*(blackbody + Comptb + Gaussian)$, where $N_H$ is fixed at a value  3.00$\times 10^{21}$ cm$^{-2}$ \citep{Bloser00}; 
color temperature $T_s$ and $T_{BB}$ are fixed at 1.3 and 0.7 keV, respectively (see comments in the text); 
$^{\dagger\dagger}$ when parameter $\log(A)\gg1$, this parameter is fixed at 2.0 (see comments in the text), 
$^{\dagger\dagger\dagger}$ normalization parameters of $blackbody$ and $Comptb$ components are in units of 
$10^{-2}\times L_{39}/d^2_{10}$ $erg/s/kpc^2$, where $L_{39}$ is the source luminosity in units of 10$^{39}$ erg/s, 
$d^2_{10}$ is the distance to the source in units of 10 kpc 
and $Gaussian$ component is in units of $10^{-2}\times total~~photons$ $cm^{-2}s^{-1}$ in line, 
$^{\dagger\dagger\dagger\dagger}$spectral fluxes (F$_1$/F$_2$) in units of $\times 10^{-9}$ ergs/s/cm$^2$ for  (3 -- 10) and (10 -- 60) keV energy ranges respectively.  
\end{deluxetable}
~~~~

\begin{deluxetable}{lccccccccccc}
\rotate
\tablewidth{0in}
\tabletypesize{\scriptsize}
    \tablecaption{Comparison of parameters of atoll sources 4U~1820-30, GX~3+1$^1$ and 4U~1728-34$^2$}
    \renewcommand{\arraystretch}{1.2}
\tablehead{Source&  D, &Type of &Mass of   & $P_{orb}$,   &    $i$,    &$kT_e$ &$N_{comptb}$,           &$kT_{BB}$,&$kT_s$,& $f$&$t_{LS-HS-LS}$, \\
                 & kpc & donor  &donor star&    min       & degrees    &  keV  &$L^{soft}_{39}/D^2_{10}$& keV      & keV   &           & days    \\
                 &     & star   &M$_\odot$ &              &            &       &                        &          &       &           &         }
 \startdata
GX~3+1           &4.5$^3$  &   A    &$\sim$ 10&  -           & -          &2.3-4.5& 0.04-0.15          & 0.6      & 1.16-1.7& 0.2-0.9  & 1000      \\
4U~1820-30       &5.8-8$^4$&   WD   &  0.07   &11.46$\pm$0.04&43$\pm$9$^5$&2.9-21.0& 0.02-0.14         & 0.6      & 1.1-1.7 & 0.2-1    & 170       \\
4U~1728-34       &4.2-6.4$^6$&  ?    &   -     &  -           & -          &2.5-15 & 0.02-0.09          & 0.6-0.7  & 1.3   & 0.5-1      & 15        \\
      \hline
      \enddata
    \label{tab:par_bbody}
References:
(1) ST12; 
(2) ST11;
(3) Kuulkers \& van der Klis  (2000); 
(4) Shaposhnikob \& Titarchuk (2004);
(5) Arons \& King (1993); 
(6) van Paradijs (1978)
\end{deluxetable}

\newpage              

\newpage 

%
%

\newpage

%
%

\newpage

%
%

\newpage

\newpage

%
%

\newpage

%
%

\newpage
\begin{figure}[ptbptbptb]
\includegraphics[scale=0.8,angle=0]{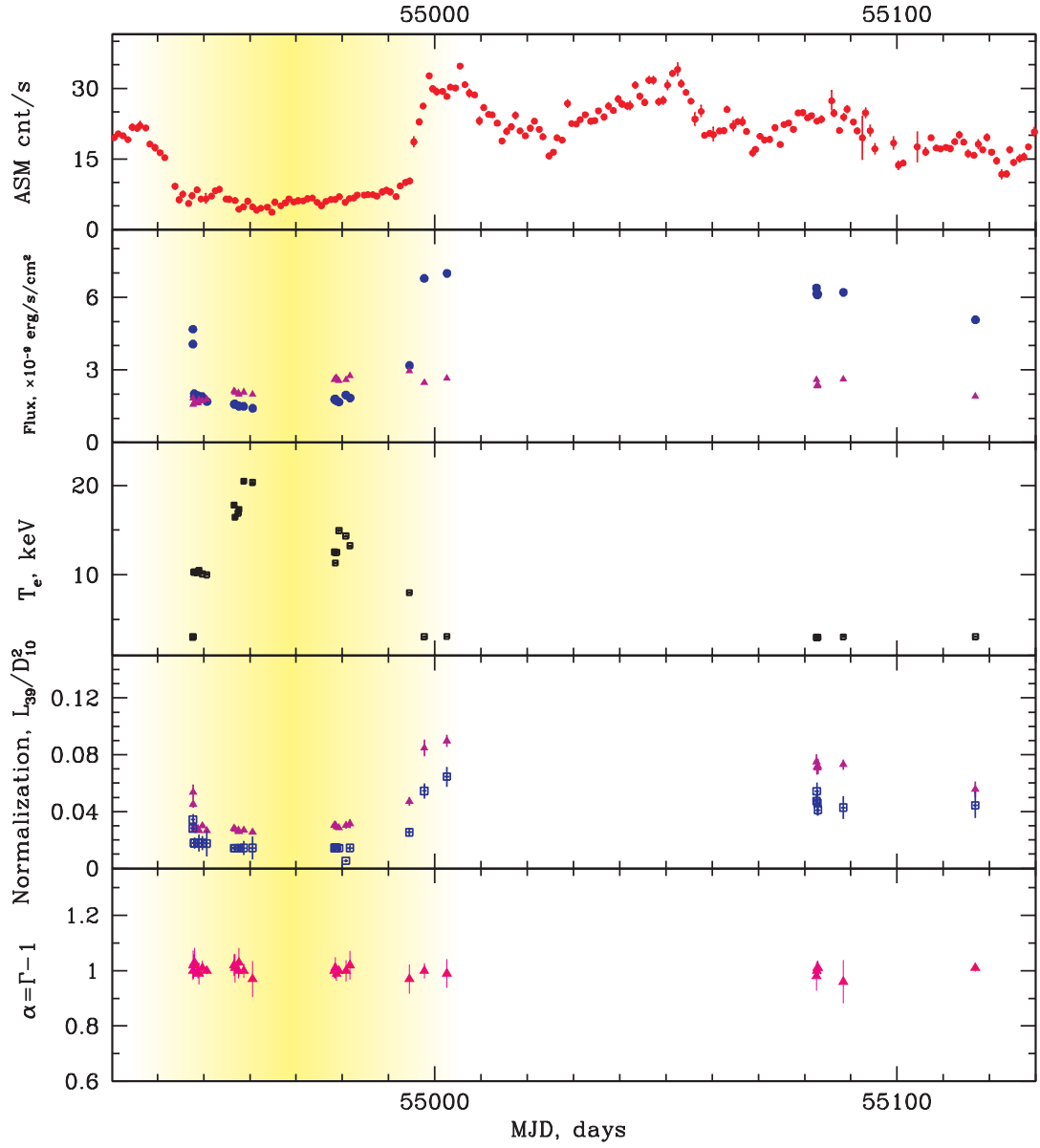}
\caption{ Similar to  that presented in Fig. \ref{lc_1996} but  for the {\it RXTE} 2009 
transition set 
  {\it R7}.  The {\it quasi-plateau} phases of the low luminosity state of 4U~1820-30 are marked using orange vertical strip. 
}
\label{evolution_low}
\end{figure}

%
%

%
%

\newpage
\begin{figure}[ptbptbptb]
\includegraphics[scale=1.0,angle=0]{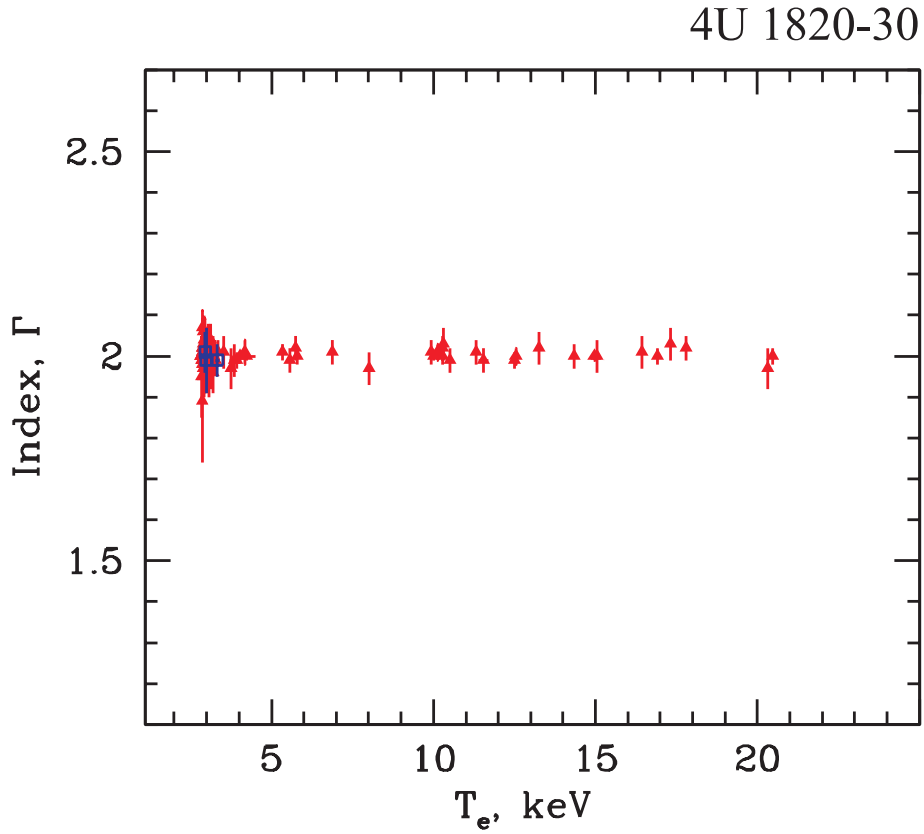}
\caption{
The photon index $\Gamma$ plotted versus the electron temperature $kT_e$ (in keV)  in the frame of  our spectral model 
$wabs*(blackbody+Comptb+Gaussian)$ during 
transition events (see Tables 3, 4). 
 Blue and   red points correspond to {\it Beppo}SAX  and {\it RXTE} observations.
}
\label{gam_Te}
\end{figure}

\newpage
\begin{figure}[ptbptbptb]
\includegraphics[scale=0.7,angle=0]{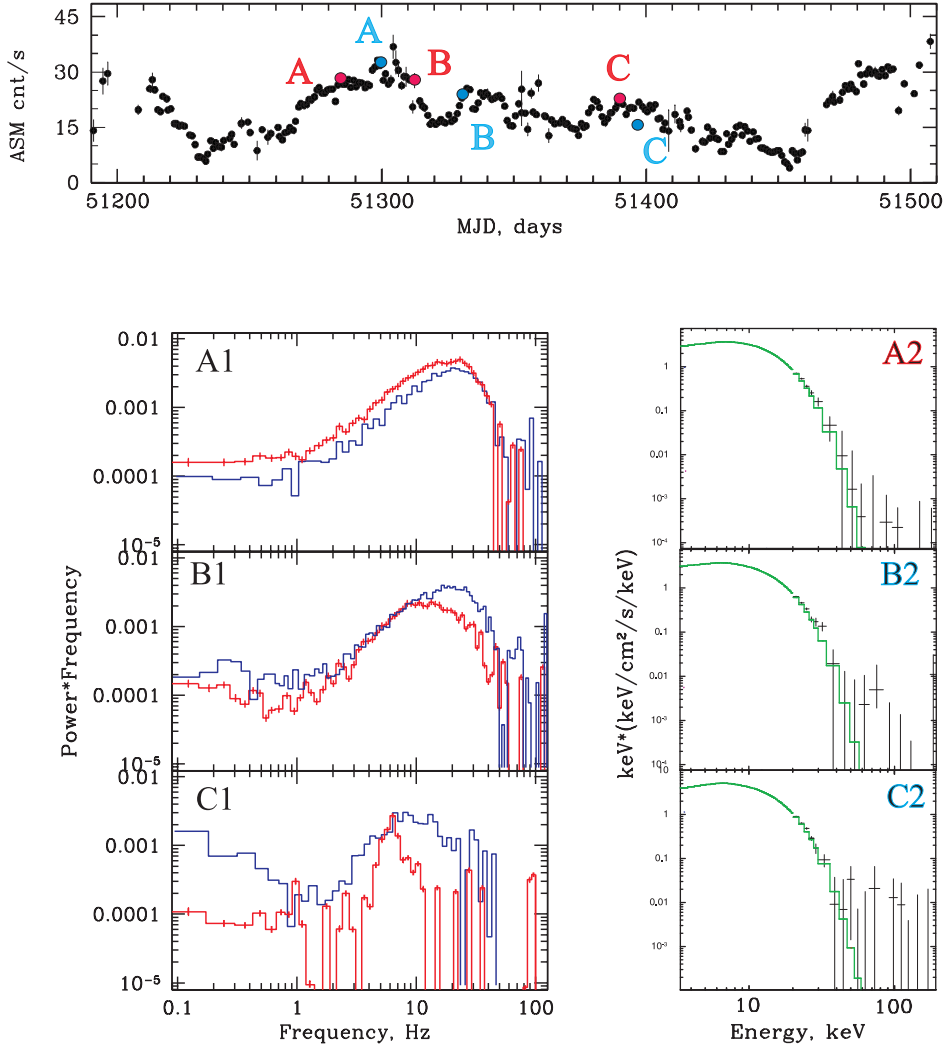}
\caption{
{\it Top}: evolution of {\it RXTE}/ASM count rate during the high luminosity state at $R3$ (1999) transition events. Red/blue points 
A, B, and C mark moments at MJD = 51283.6/51300, 51313.7/51330.5 and 51389.4/51396.26 covering different 
transition phases.  
$Bottom$: 
PDSs for 15-30 keV  band ($left$ column) are plotted along with the $E*F(E)-$diagram ($right$ column) 
for A ($red$, top), B ($blue$, middle) and C ($blue$, bottom) points  of X-ray light curve.  
All points are related to  the $banana$ state [strong broadband noise, VLFN and QPOs at $\nu_l\sim$6 -- 7 Hz (C $red$)]. 
The $E*F(E)-$diagrams (panels A2, B2, C2) are related to the corresponding power spectra 
(panels A1, B1, C1).  The data are shown by black points.
The electron temperature $kT_e$ of the corresponding 
energy spectra of 4U~1820-30 is about  3 keV.
}
\label{ev_PDS_SP_high}
\end{figure}

%
%

\newpage
\begin{figure}[ptbptbptb]
\includegraphics[scale=0.7,angle=0]{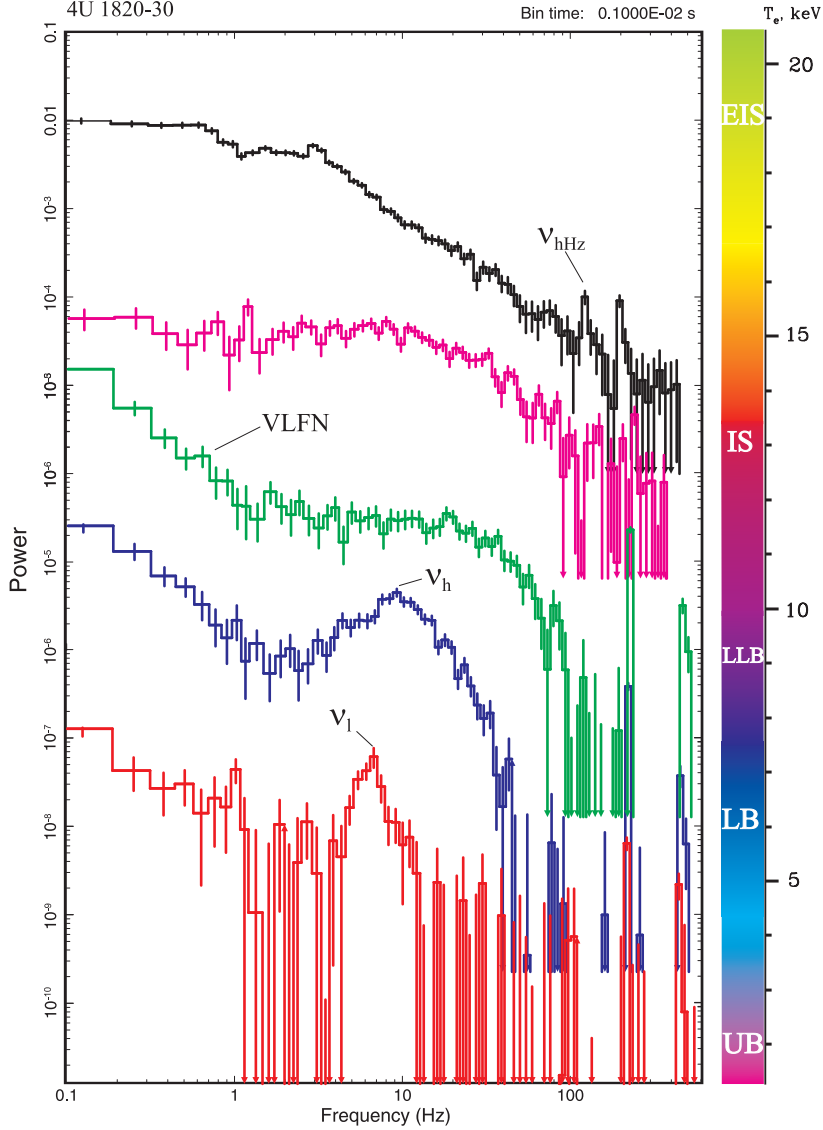}
\caption{
PDSs of 4U~1820-30 related to its X-ray spectral states.
$kT_e$ values (in keV)  corresponding  to the energy spectra are indicated at the right 
vertical axis. PDSs in the extreme island state (EIS), island state (IS, multiplied by factor $10^{-2}$ for clarity),
lower left banana state (LLB, $\times 10^{-4}$), lower banana state (LB, $\times 10^{-6}$) and upper banana state (UB, $\times 10^{-8}$) 
are presented from $top$ to $bottom$. The histograms
consist of  three components: VLFN ({\it very low frequency noise} in $banana$ states), the peaked 
noise component,  
low-frequency QPOs are  fit by
Lorentzians ($\nu_{l}$, $\nu_{h}$) and high frequency QPO ($\nu_{hHz})$.
}
\label{PDS}
\end{figure}

%
%

\newpage
\begin{figure}[ptbptbptb]
\includegraphics[scale=0.7,angle=0]{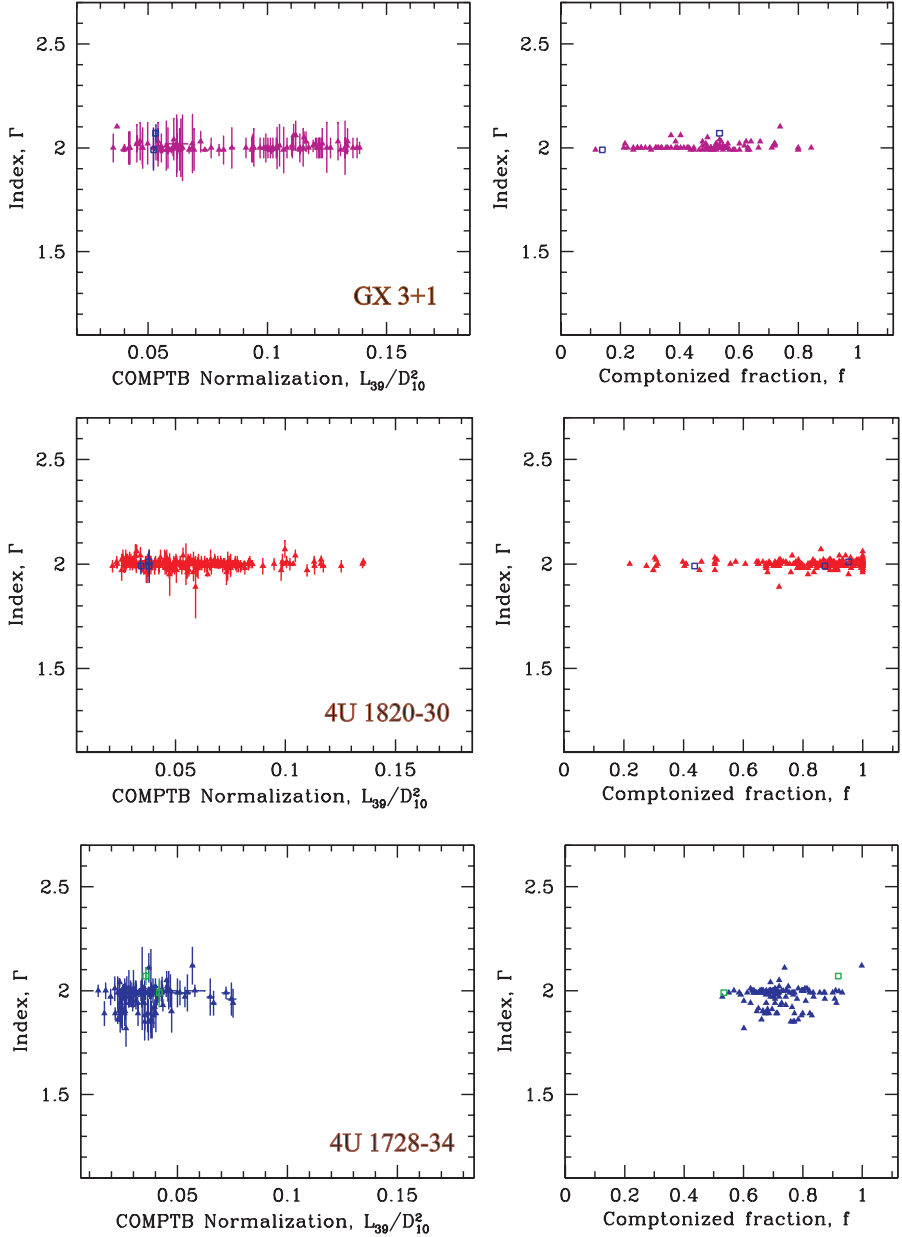}
\caption{
{\it From Top to Bottom:} 
Plots $\Gamma$ vs Comptb Normalization ({\it left column}) and vs Comptonized fraction $f$ ({\it right column}) 
for  GX~3+1 ($top$), 4U~1820-30 ($middle$) and 4U~1728-34 ($bottom$) 
obtained using the  $wabs*(blackbody+Comptb+Gaussian)$  model. 
On the $top$ panels  $crimson$ and $blue$  points are for GX~3+1 taken from ST12 and  on the  middle panels   for 4U 1820-30 $red$ and  $blue$ points  correspond to {\it RXTE}  and  $Beppo$SAX data respectively (current study). On the $bottom$ panels $blue$  and $green$  points correspond to {\it RXTE}  and $Beppo$SAX data respectively  for 4U 1728-34  (data  taken from ST11). 
}
\label{gam_norm_3obj}
\end{figure}

%
%

\newpage

\begin{figure}[ptbptbptb]
\includegraphics[scale=0.98,angle=0]{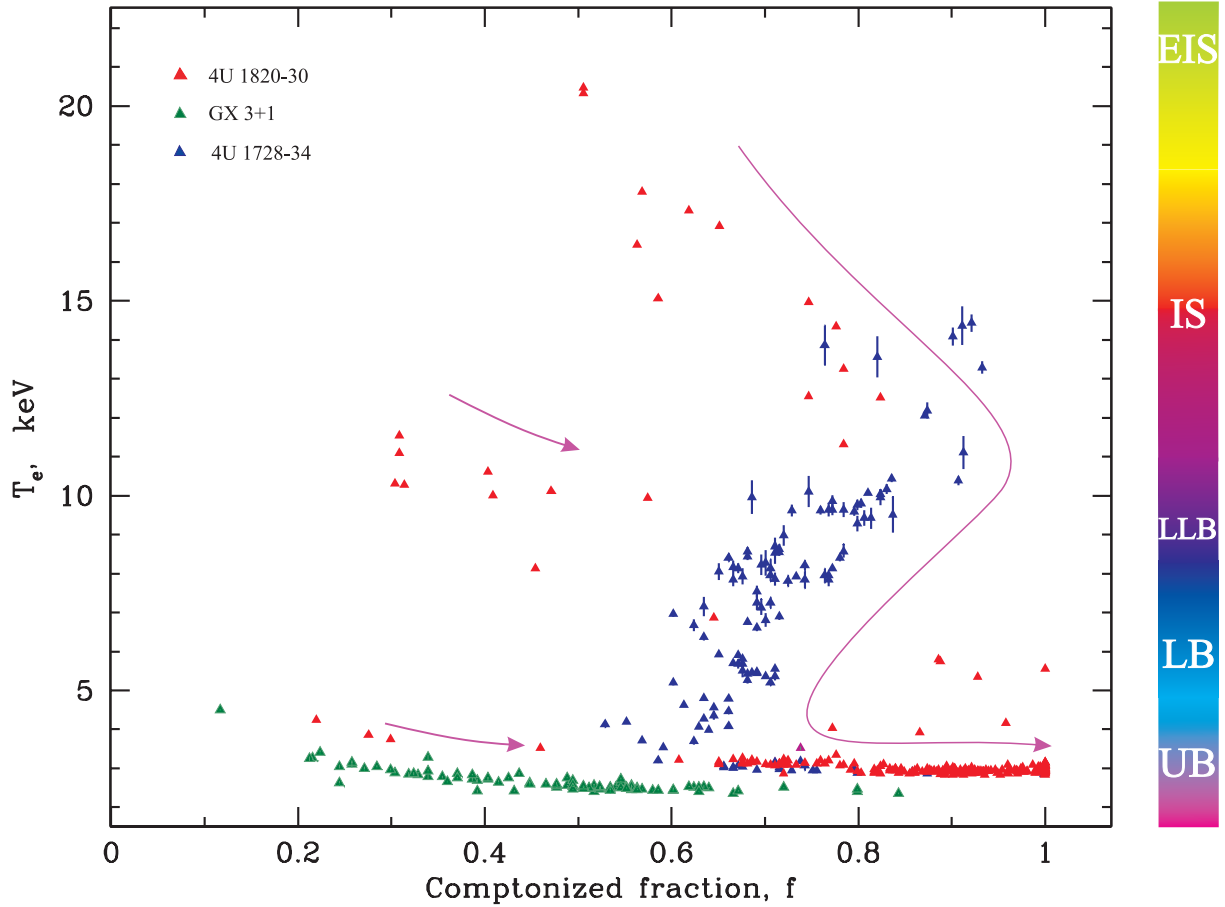}
\caption{
$kT_e$ (in keV)  plotted versus illumination fraction $f$ for  4U~1820-30, 
 GX~3+1 (taken from ST12) and 4U~1728-34 (taken from ST11) during spectral state transitions 
obtained using  the  $wabs*(blackbody+Comptb+Gaussian)$ model. {\it Red, green}  and {\it blue} 
points correspond to {\it RXTE} 
observations of 4U~1820-30, GX~3+1 and 4U~1728-34 respectively. 
The {\it bended arrows} are related to   an increase of mass accretion rate.
On the right-hand side of  the Figure we show  a sequence of CCD states 
(EIS -- extreme island state, 
IS --  island state,
LLB -- lower left banana state,
LB -- lower banana state and 
UB -- upper banana state) which  are listed according to the standard {\it atoll}$-Z$ scheme~\citep{hasinger89}.
One can see that  
$kT_e$  is directly related to the sequence of CCD stages.
}
\label{T_e_vs_f_comp}
\end{figure}

%
%



%
%

\newpage
\begin{figure}[ptbptbptb]
\includegraphics[scale=0.90,angle=0]{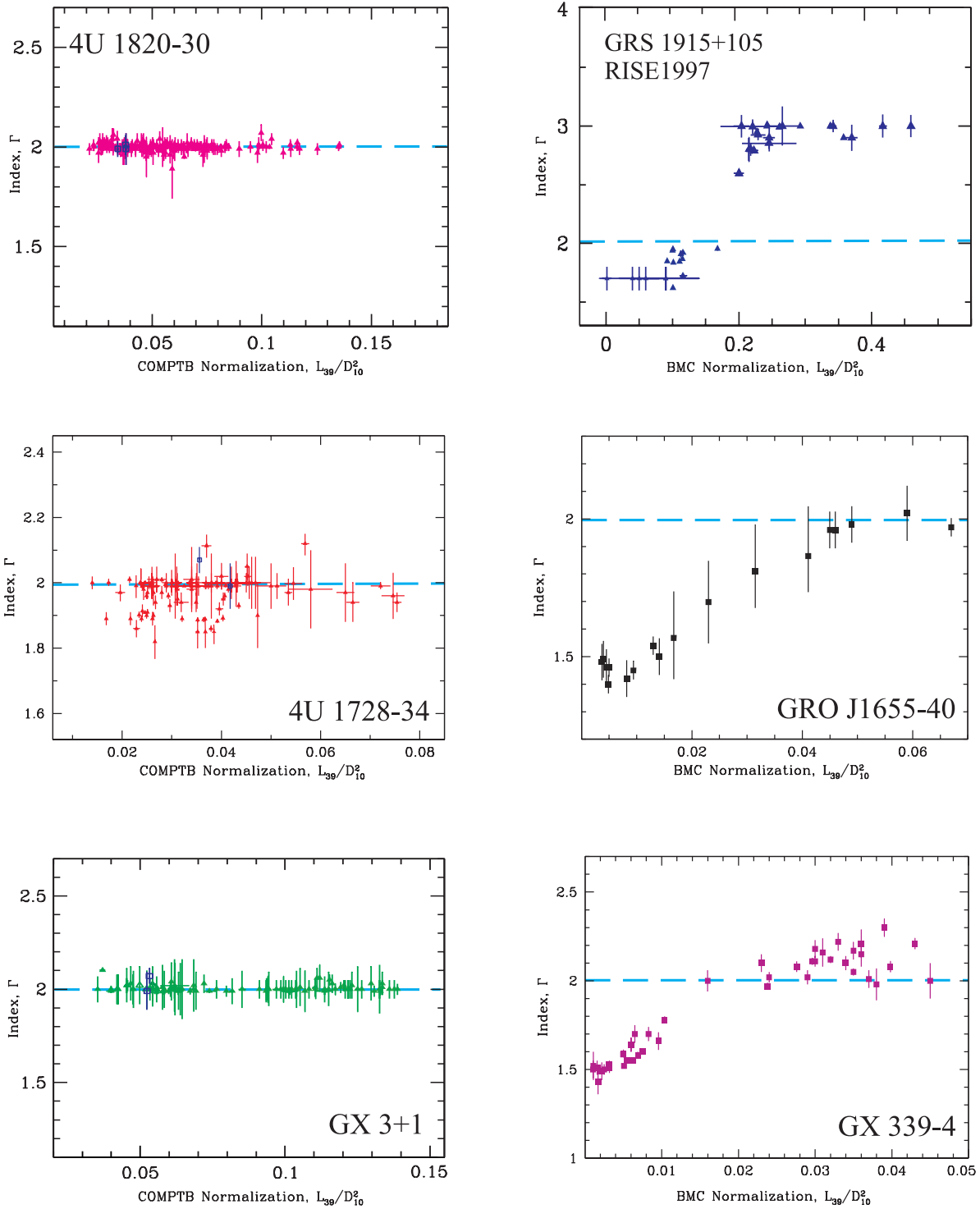}
\caption{Examples of diagrams of the photon index $\Gamma$ versus 
 the COMPTB normalization (proportional to mass accretion rate)  for BH sources 
[right column, GRS~1915+105 (TS09), GRO~J1655-40 (ST08) and GX~339-4 (ST08)] along with 
{\it atoll} 
NS sources [left column, 4U~1820-30, 4U~1728-34 (ST11) and GX~3+1 (ST12)]. For  all plots the {\it RXTE} data were 
used along with $Beppo$SAX data (indicated by blue points on the $left$ column). 
One can see a noticeable change of $\Gamma$ followed by  the saturation plateau for BHs as for NSs   
the index only  slightly varies about 2. 
The level for $\Gamma = 2$ is indicated by  blue dashed line.
}
\label{bh_ns_examples}
\end{figure}



\begin{thebibliography}{}



\bibitem[Asai et al. (1994)]{asai}
Asai, K., Dotani, T., Mitsuda, K., Nagase, F., Kamado, Y., Kuulkers, E., \& Breedon, L. M. 1994, PASJ, 
46, 479

\bibitem[Bloser et al. (2000)]{Bloser00}
Bloser, P.~F., Grindlay, J. E., Kaaret, P. et al. 2000, ApJ, 542, 1000

\bibitem[Bloser et al. (1996)]{Bloser96}
Bloser, P. F., et al. 1996, A\&AS, 120, 275

\bibitem[Boella et al. (1997)]{boel97} 
Boella,  G. et al.  1997, A\&AS, 122, 327

\bibitem[Bradt et al. (1993)]{bradt93} 
Bradt,  H.V., Rothschild, R.E. \& Swank, J.H. 1993, A\&AS, 97, 355

\bibitem[Clark et al. (1977)]{clark77} 
Clark, G.W., Li, F.K., Canizares, C., Haykava, S., Jernigan, G., Lewin, W.H.G., 1977, MNRAS, 179, 651

\bibitem[Cornelisse et al. (2003)]{Cornelisse03}
Cornelisse, R., et al. 2003, A\&A, 405, 1033

\bibitem[Costantini et al. (2012)]{Costantini11} 
Costantini, E.\ et al.  2012, A\&A, 539, 32


\bibitem[Chou \& Grindlay (2001)]{Chou01}
Chou, Y.  \&  Grindlay, J. E. 2001, ApJ, 563, 934

\bibitem[Christian, D. J., \& Swank (1997)]{Christ97}
Christian, D. J., \& Swank, J. H. 1997, ApJS, 109, 177

\bibitem[D Ai  et al. (2006)]{dai2006} 
D' Ai, A. et al. 2006, A\&A, 448, 817

\bibitem[D'Amico et al. (2001)]{amico}
D'Amico, F., Heindl, W. A., Rothschild, R. E., \&  Gruber, D. E. 2001, ApJL, 547, L147


\bibitem[Di Salvo  et al. (2000)]{Di_Salvo2000} 
Di Salvo, T., Iaria, R., Burderi, L., \& Robba, N. R. 2000, ApJ, 542, 1034

\bibitem[Ebisawa  et al. (1994)]{Ebisawa94} 
Ebisawa, K., et al. 1994, PASJ, 46, 375

\bibitem[Egron et al. (2001]{ergon11}
Egron, E., Di Salvo, T., Burderi, L., et al. 2011, A\&A, 530, A99

\bibitem[Hirano et al. (1987]{hirano87}
Hirano, T., Hayakawa, S., Nagase, F., Masai, K., \& Mitsuda, K. 1987,
PASJ, 39, 619

\bibitem[Farinelli \& Titarchuk (2011)]{ft11}
Farinelli,~R. \& Titarchuk,~L., 2011, A\&A,   525, 102 (FT11)

\bibitem[Farinelli et al. (2008)]{F08}
Farinelli,~R., Titarchuk,~L., Paizis, A. \& Frontera, F. 2008, \apj, 680, 602, (F08)

\bibitem[Farinelli et al. (2005)]{Far05}
Farinelli, R., Frontera, F., Zdziarski, A. A., Stella, L., Zhang, S. N., van der Klis, M., 
Masetti, N., \& Amati L. 2005, A\&A, 434, 25

\bibitem[Ford et al. (2000)]{F0rd2000}
Ford, E. C., van der Klis, M., Mendez, M., et al. 2000, ApJ, 537, 368

\bibitem[Frontera et al. (1997)]{fron97}
Frontera, F. et al.  1997, SPIE, 3114, 206

\bibitem[Geldzahler (1983)]{Geldzahl83}
Geldzahler, B. J., 1983, ApJ, 264, L49

\bibitem[Gierli'nski et al. (1999)]{gerl}
Gierli'nski, M., Zdziarski, A. A., Poutanen, J., Coppi, P. S., Ebisawa, K., \& Johnson, W. N. 1999, MNRAS, 309, 496

\bibitem[Grindlay et al. (1976)]{Grindlay76}
Grindlay, J., Gursky, H., Schnopper, H., Parsignault, D. R., Heise, J., Brinkman,
A. C., \& Schrijver, J. 1976, ApJ, 205, L127

\bibitem[Grindlay \& Seaquist (1986)]{Grindlay86}
Grindlay, J.E. \& Seaquist, E.R. 1986, ApJ, 310, 172

\bibitem[Hasinger \& van der Klis (1989)]{hasinger89}
Hasinger, G. \& van der Klis, M. 1989, A\&A, 225, 79

\bibitem[Hirano et al. (1987)]{hirano87}
Hirano, T.  Haykawa, S., Nagase, F. Masai, K. \& Mitsuda,  K. 1987, PASJ, 39, 619

\bibitem[Kaaret et al. (1999)]{kaaret99}
Kaaret, P., Piraino, S., Bloser, P. F., Ford, E. C., Grindlay, J. E., Santangelo, A., Smale, A. P., \& Zhang, W. 1999, ApJ, 520, L37

\bibitem[Krimm et al. (2009)]{Krimm09}
Krimm, H. A. et al., 2009, ATel N 2071

\bibitem[Kuulkers et al. (2003)]{kuulk03}
Kuulkers, E., den Hartog, P. R., in't Zand, J. J. M., Verbunt, F. W. M., Harris,
W. E., \& Cocchi, M., 2003, A\&A, 399, 663

\bibitem[Kuulkers \& van der Klis  (2000)]{kk00}
Kuulkers, E.  \& van der Klis, M. 2000, A\&A, 356, L45

\bibitem[Ku$\acute s$mierek et al. (2011)]{kusmierek11} 
Ku$\acute s$mierek, K., Madej, J. \& Kuulker, E. 2011, MNRAS,  
in press (arXiv: 1105.1525v)

\bibitem[Lewin et al. (1993)]{Lewin93}
Lewin, W.H.G., van Paradijs, J., \& Taam, R.E. 1993, Space Sci. Rev., 62, 223

\bibitem[Lin et al. (2007)]{LRH07}
Lin, D., Remillard, R., \& Homan, J. 2007, Apj, 667, 1073 (LRH07)

\bibitem[McConnell et al., 2002]{mcconel}
McConnell, M. L., et al. 2002, ApJ, 572, 984

\bibitem[Marti et al., 1998]{marti98}
Marti, J., Mirabel, I.F., Rodriguez, L.F., \& Chaty, S. 1998, A\&A, 332, L45

\bibitem[Migliari et al. (2004)]{Migliari04}
Migliari, S, Fender, R.P., Rupen, M. et al. 2004, MNRAS, 351, 186

\bibitem[Morrison \& McCammon (1983)]{morr83}
Morrison, R. \& McCammon, D. 1983, ApJ 270, 119

\bibitem[Ng et al. (2010]{Ng10}
Ng, C., Daz, T. M., Cadolle, B. M., \& Migliari, S. 2010, A\&A, 522, A96

\bibitem[Oosterbroek  et al. (2001)]{ooster2001} 
Oosterbroek, T., Barret, D., Gianazzi, M., \& Ford, E. C. 2001, A\&A, 366, 138

\bibitem[Parmar et al. (1997)]{parmar97} 
Parmar, A. N., et al.  1997, A\&AS, 122, 309

\bibitem[Parsignault \& Grindlay (1978)]{parsignault78} 
Parsignault, D. R., \& Grindlay, J. E. 1978, ApJ, 225, 970

\bibitem[Piraino et al (2000)]{Piraino00}
Piraino, S., Santangelo, A., \& Kaaret, P. 2000, A\&A, 360, 35

\bibitem[Piraino et al (1999)]{Piraino99}
Piraino, S., Santangelo, A., Ford, E. C., \& Kaaret, P. 1999, A\&A, 349, L77 

\bibitem[Priedhorsky \& Terrell (1984)]{PT84}
Priedhorsky, W. \& Terrell, J., 1984, ApJ, 284, L17

\bibitem[Rappaport et al. (1987)]{rap87}
Rappaport, S., Nelson, L. A., Ma, C. P., \& Joss, P. C. 1987, ApJ, 322, 842

\bibitem[Rich et al. (1993)]{Rich93}
Rich, R. M., Minniti, S., \& Liebert, J. 1993, ApJ, 406, 489

\bibitem[Seifina \& Titarchuk  (2012)]{st12}   
Seifina, E. \& Titarchuk, L. 2012,  \apj, 747, 99   (ST12)

\bibitem[Seifina et al.   (2013)]{st13}   
Seifina, E.,  Titarchuk, L. \& Frontera, F.  2013,  \apj,  accepted

\bibitem[Seifina \& Titarchuk  (2011)]{st11}   
Seifina, E. \& Titarchuk, L. 2011,  \apj, 737, 128   (ST11)

\bibitem[Seifina \& Titarchuk  (2010)]{ST10}   
Seifina, E. \& Titarchuk, L. 2010,  \apj, 722, 586 (ST10)





\bibitem[Sidoli et al. (2001)]{Sidoli01}
Sidoli, L., Parmar, A. N., Oosterbroek, T., Stella, L., Verbunt, F., Masetti, N., \& Dal Fiume, D. 2001, A\&A, 368, 451

\bibitem[Simon (2003)]{Simon03}
Simon, V. 2003 , A\&A, 405, 199

\bibitem[Shakura \& Sunyaev  (1973)]{ss73} Shakura, N.~I., \& Sunyaev, R.~A. 1973, \aap, 24, 337  

\bibitem[Shaposhnikov \& Titarchuk (2009)]{st09}
Shaposhnikov, N., \& Titarchuk, L. 2009, ApJ, 699, 453

\bibitem[Shaposhnikov \& Titarchuk (2004)]{ST04}
Shaposhnikov, N., \& Titarchuk, L. 2004, ApJ, 606, L57

\bibitem[Smale et al. (1994)]{smale94}
Smale, A. P., Dotani, T., Mitsuda, K., \& Zylstra, G. 1994, BAAS, 26, 872

\bibitem[Smale et al. (1997)]{smale97}
Smale, A. P., Zhang, W., \& White, N. E. 1997, ApJ, 483, L119


\bibitem[Stella et al. (1987)]{stel87}
Stella, L., White, N. E., \& Priedhorsky, W. 1987, ApJ, 315, L49


\bibitem[Strohmayer, T. \& Bildsten (2004)]{strohm04}
Strohmayer, T. \& Bildsten, L., 2004, "New view of thermonuclear bursts", in
Compact Stellar X-Ray Sources, Cambridge Astrophys. Ser. 26, eds. W.H.G.
Lewin and M. van der Klis

\bibitem[Strohmayer \& Brown (2002)]{strohm02}
Strohmayer, T. E. \& Brown, E. F. 2002, ApJ, 566, 1045

\bibitem[Sunyaev \& Titarchuk (1980)]{ST80}
Sunyaev, R. A. \& Titarchuk, L. G. 1980, A\&A, 86, 121 

\bibitem[Tarana et al. (2006)]{tarana06}
Tarana, A., Bassano, A., Ubertini, P. and Zdziarski, A. A. 2006, ApJ, 654, 494 

\bibitem[Titarchuk et al.  (1998)]{tlm98}   Titarchuk, L., Lapidus, I.I. \& Muslimov, A.   1998,
 \apj, 499, 315

\bibitem[Titarchuk,  Seifina \& Frontera (2013)]{tsf13}   
Titarchuk, L.  Seifina, E. \& Frontera, F.   2013, \apj, 767,  (TSF13)

\bibitem[Titarchuk \& Seifina  (2009)]{tsei09}   
Titarchuk, L. \& Seifina, E.   2009, \apj, 706, 1463

\bibitem[Vacca et al. (1986)]{vacca86}
Vacca, W.~D., Lewin, W.~H.~G. \& Paradijs, J. 1986, MNRAS, 220, 339

\bibitem[van Paradijs  (1978)]{par78} 
van Paradijs, J. 1978,  Nature, 274, 650

Wardzi'nski, G., Zdziarski, A. A., Gierli'nski, M., Eric Grove, J., Jahoda, K., \& Neil Johnson, W. 2002, 
MNRAS, 337, 829

\bibitem[Wardzi'nski et al. (2002)]{wardz}
Wardzi'nski, G., Zdziarski, A. A., Gierli'nski, M.,  Grove, J.E., Jahoda, K., \& Neil Johnson, W. 2002, 
MNRAS, 337, 829

\bibitem[Wen et al. (2006)]{Wen06}
Wen, L., Levine, A. M., Corbet, R. H. D., \& Bradt, H. V. 2006, ApJS, 163, 372

\bibitem[Wijnands \& van der Klis (1999)]{Wijnand99}
Wijnands, R. \& van der Klis, M., 1999, ApJ 514, 939

\bibitem[White et al. (1986)]{White86}
White, N. E., Peacock, A., Hasinger, G. et al. 1986, MNRAS 218, 129

\bibitem[Zdziarski et al. (2007)]{Zdziarski07}
Zdziarski, A. A., Wen, L., \& Gierlinski, M. 2007, MNRAS, 377, 1006

\bibitem[Zhang et al. (1998)]{Zhang98}
Zhang, W., Smale, A. P., Strohmayer, T. E., \& Swank, J. H. 1998, ApJ, 500, L171


\end{thebibliography}
\end{document}